\documentclass[12pt]{article}

\newcommand{\bbr}{I\!\! R}
\newcommand{\bbz}{Z\!\!\! Z}

\newcommand{\x}{arXiv:}
\newcommand{\m}{\mathrm}

\usepackage{graphicx}

\begin{document}
\thispagestyle{empty}
\begin{center}

\null \vskip-1truecm \vskip2truecm {\Large{\bf

\textsf{Singularities in and Stability of Ooguri-Vafa-Verlinde
Cosmologies}

}}

\vskip1truecm {\large \textsf{Brett McInnes}} \vskip1truecm

 \textsf{Abdus Salam ICTP \\ and \\  National
  University of Singapore\footnote{Permanent address}}

email: matmcinn@nus.edu.sg\\

\end{center}
\vskip1truecm \centerline{\textsf{ABSTRACT}} \baselineskip=15pt
\medskip

Ooguri, Vafa, and Verlinde have recently proposed an approach to
string cosmology which is based on the idea that cosmological
string moduli should be selected by a Hartle-Hawking wave
function. They are led to consider a certain Euclidean space which
has \emph{two different Lorentzian interpretations}, one of which
is a model of an \emph{accelerating cosmology}. We describe in
detail how to implement this idea without resorting to a ``complex
metric". We show that the four-dimensional version of the OVV
cosmology is null geodesically incomplete but has no curvature
singularity; also that it is [barely] stable against the
Seiberg-Witten process [nucleation of brane pairs]. The
introduction of matter satisfying the Null Energy Condition has
the paradoxical effect of both stabilizing the spacetime and
rendering it genuinely singular. We show however that it is
possible to arrange for an \emph{effective} violation of the NEC
in such a way that the singularity is avoided and yet the
spacetime remains stable. The possible implications for the early
history of these cosmologies are discussed.

 \vskip3.5truecm
\begin{center}

\end{center}

\newpage

\addtocounter{section}{1}
\section* {\large{\textsf{1. The Perils of Ooguri-Vafa-Verlinde Cosmologies}}}
Ooguri, Vafa, and Verlinde have put forward \cite{kn:OVV} an
approach to string cosmology which implements the ``wave function
of the Universe" programme of Hartle and Hawking \cite{kn:hartle}.
They do not propose a specific four-dimensional cosmological
model, but here we shall see that much can be said about ``OVV
cosmology" [see also \cite{kn:dijk}] even at this early stage.

Ooguri et al begin by emphasising that it is natural, in the
context of quantum gravity, to assume that the spatial sections of
our universe are compact, a point also recently stressed by Linde
\cite{kn:lindetypical}\cite{kn:lindenew} from a different point of
view. This assumption plays a crucial role in the Ooguri et al
proposal, since it means that various string moduli are not fixed
--- they must be selected by a wave function with amplitudes
peaked at the appropriate moduli values.

Ooguri et al work with a compactification to Euclidean
two-dimensional anti-de Sitter space, the hyperbolic space H$^2$;
they use the foliation of H$^{\m{n}}$, familiar from studies of
the AdS/CFT correspondence, by \emph{flat} slices \cite{kn:oz}. In
the case of H$^2$, they use this foliation to perform a further
compactification of H$^2$ to a space with topology
$\bbr\,\times\,\m{S}^1$, where S$^1$ is a circle --- or, rather, a
one-dimensional torus. The discussion in \cite{kn:OVV} is based
entirely on this two-dimensional space and the \emph{two
different} Lorentzian  spacetimes which Ooguri et al derive from
it.

Ultimately, of course, Ooguri et al hope to obtain a
quasi-realistic \emph{four-dimensional} cosmology in this way. We
argue that, whatever the precise form of this spacetime ultimately
proves to be, one should expect the topology of its spatial
sections to be that of a \emph{torus} [or a quotient of a torus].
There are several reasons for this. First, it would allow us to
make contact with the work of Linde
\cite{kn:lindetypical}\cite{kn:lindenew} mentioned above; second,
it is compatible with the well-known Brandenberger-Vafa string gas
cosmologies \cite{kn:brandvafa}; third, granted compactness, the
observed \cite{kn:spergel} near-flatness of our Universe also
points to toral spatial sections. Finally, toral spatial sections,
if they are of the right size, allow us to answer many of the
objections which have been raised against the existence of
horizons in accelerating cosmologies. Note that the size of the
tori is not determined by the spacetime curvature, so it must be
fixed in some other way\footnote{By contrast, the radius of the
smallest spherical spatial slice of the simply connected global de
Sitter spacetime is completely fixed by the cosmological
constant.}; one can hope that it is determined by the wave
function of the Universe. The fundamental importance of this
spatial size ``modulus" will appear at several points in our work.
[See \cite{kn:reallyflat} for an extended discussion of various
aspects of spacetimes with flat, compact spatial sections.]

The simplest way to proceed towards four dimensions is as follows.
The standard foliation of H$^4$ by flat sections can be used to
define a partial compactification of H$^4$ to a space with
topology $\bbr\,\times\,\m{T}^3$, where T$^3$ is the three-torus.
This is exactly analogous to the OVV procedure, and so we shall
assume that four-dimensional space and spacetime have this
structure. This simple assumption has far-reaching consequences,
however.

A recent development in string theory has been the realization
that as one moves away from exact Euclidean AdS [that is,
hyperbolic, H$^{\m{n}}$] geometry
--- as of course one ultimately must \cite{kn:wittenads} --- various
novel effects arise, and eventually one reaches geometries where
specifically stringy effects \cite{kn:porrati} destabilize a
spacetime \emph{which may appear perfectly acceptable from a
classical point of view}.

For a concrete example, parametrise the three-torus by angles
[taking a common minimal radius of 2K for the circles] and endow
$\bbr\;\times\;\m{T}^3$ with the Euclidean metric
\begin{equation}\label{eq:A}
g\m{_c(2,\,2K,\,L)_{++++} \;=\; +\,dt^2\; +\;
4K^2\;cosh^2({{t}\over{L}})[d\theta_1^2 \;+\; d\theta_2^2 \;+\;
d\theta_3^2]},
\end{equation}
where L is a constant; here, apart from the indication of the
signature and the subscript c, which reminds us of the cosh
function, the notation is as in \cite{kn:unstable}\footnote{If
this is a metric for a string gas cosmology, or if one wants to
use it in the Hartle-Hawking manner, one should restrict t to t
$\geq$ 0 in this formula.}. Notice that, ignoring the signature,
the ``scale factor" is precisely that of de Sitter spacetime ---
only the geometry and topology of the spatial sections is
different. The geometry has no obvious pathology; the manifold is
non-singular and inextensible. For large t, we have
\begin{equation}\label{eq:B}
g\m{_c(2,\,2K,\,L)_{++++} \;\approx\; +\,dt^2\; +\;
K^2\,e^{2\,t/L}\,[d\theta_1^2 \;+\; d\theta_2^2 \;+\;
d\theta_3^2]}.
\end{equation}
The right side of this relation is the metric of constant
curvature $-\,1$/L$^2$ on a partial compactification of H$^4$; it
is precisely a four-dimensional version of the metric on the
partially compactified two-dimensional hyperbolic space discussed
by Ooguri et al \cite{kn:OVV}. Thus we can think of the geometry
corresponding to $g_{\m{c}}(2,\,\m{2K,\,L)_{++++}}$ as being
obtained by deforming a certain quotient of ordinary Euclidean AdS
space. Near infinity, the two spaces are almost indistinguishable
geometrically. Physically, however, they are very different, as we
shall see.

The difference is revealed by the behaviour of branes on these
spaces. Let us introduce a 4-form field on these backgrounds, and
consider the nucleation of Euclidean BPS 2-branes
\cite{kn:seiberg}. The stability of this system is determined by a
\emph{purely geometric} question: can the area of a brane always
grow quickly enough to keep the action positive? In ordinary
[non-compactified] Euclidean AdS it does, but in certain distorted
versions of AdS it does not; thus we obtain a criterion for the
stringy stability of the Euclidean version of a given spacetime.
This Seiberg-Witten \cite{kn:seiberg} mechanism has been applied
to topologically non-trivial black hole spacetimes in
\cite{kn:black}. It has recently been applied to \emph{cosmology}
by Maldacena and Maoz \cite{kn:maoz}, and their work has been
extended in various ways in
\cite{kn:buchel}\cite{kn:mcinnes}\cite{kn:answering}\cite{kn:maoz3}\cite{kn:porrati}\cite{kn:reallyflat}.
[Other uses of instability to remove candidate backgrounds may be
found in, for example, \cite{kn:mann}.]

For spaces [like those represented by the metrics in (\ref{eq:A})
and (\ref{eq:B}) above] which are \emph{flat} at infinity this
question of stability is particularly subtle \cite{kn:porrati};
the system can be stable or unstable, depending in a delicate way
on the precise details of the geometry. A direct computation
\cite{kn:unstable} shows that the brane action on the background
given by (\ref{eq:A}) is in fact unbounded below, while it is
always positive for the metric on the right side of (\ref{eq:B}):
the volume term ``wins" in the former case but not in the latter,
despite the close similarity of (\ref{eq:A}) and (\ref{eq:B}) when
t is large\footnote{That is, the approximation in (\ref{eq:B}) is
not good enough in this context.}. Thus the innocent appearance of
(\ref{eq:A}) is very misleading. This is not a consistent
background for string theory. On the other hand, the
four-dimensional OVV space with metric (\ref{eq:B}) is consistent
in this sense, despite the fact that its difference from
$g_{\m{c}}(2,\,\m{2K,\,L)_{++++}}$ decays rapidly towards
infinity. Nevertheless, the OVV space is ``close" to being
unstable: although it is locally the same as Euclidean AdS$_4$,
its global structure is different, and this affects the rate at
which the area and volume of a brane can grow.

This conclusion is somewhat disturbing, for it might well apply to
the kind of cosmological models one hopes to derive from string
theory. One does not expect de Sitter spacetime itself to suffer
from such instabilities --- but we do not live in de Sitter
spacetime: we live in a version of it that has been distorted by
the presence of matter and radiation. The danger is that this
distortion might have the same destabilizing effects as the one
which turns the stable OVV space into the unstable space with
metric (\ref{eq:A}). Notice, in particular, that our spacetime has
been distorted away from de Sitter spacetime to this extent: it
appears to be spatially flat, while de Sitter spacetime has
spherical sections. Our Universe may well, therefore, have a
conformal infinity which, as in the space with metric
(\ref{eq:A}), is flat and compact. In view of the fact that the
flat-boundary case is so delicate, \emph{an apparently realistic
string cosmology could be perilously close to being
non-perturbatively unstable.} Clarifying this question is one of
our major objectives.

We shall proceed as follows: first we discuss the unfamiliar way
in which the space considered by Ooguri et al \cite{kn:OVV}, and
its four-dimensional generalization, must be continued to
Lorentzian signature. We explain in detail the claim of Ooguri et
al that their \emph{single} Euclidean space can continue to
\emph{two} distinct Lorentzian spacetimes: this is possible with
the aid of a slight extension of Witten's \cite{kn:bubble}
technique of ``multiple analytic continuation". The continuation
can be performed in such a way that one obtains an accelerating
cosmological spacetime [``OVV spacetime"] while avoiding the
problem of a complex metric, mentioned in \cite{kn:OVV}. We then
discuss the global structure of the OVV spacetime, stressing that
it is null geodesically incomplete, though without curvature
singularities.

Next, we show that a scalar field with a certain very simple
potential can mimic the behaviour of ``conventional" matter
introduced into an initially OVV spacetime; here ``conventional"
means that the \emph{Null Energy Condition} [NEC] is satisfied.
The Einstein equations can be solved exactly when the matter
equation-of-state parameter is constant, and so we can check
directly that the brane action discussed above remains positive
throughout the corresponding Euclidean space. However, this raises
another problem, for, precisely when the NEC is satisfied, this
spacetime has to have a curvature singularity if the Einstein
equations hold, \emph{even if the usual Strong Energy Condition is
violated}. Thus we are in a paradoxical situation: the
introduction of conventional matter stabilizes the spacetime and
yet renders it singular. The only way out is to modify the
geometry of the very early Universe in some way that does not
induce Seiberg-Witten instability. This is a delicate matter,
since such modifications, if they last too long, certainly do
cause instability: indeed, this is why the space with metric
$g_{\m{c}}(2,\,\m{2K,\,L)_{++++}}$ is unstable.

The use of a scalar field to mimic the effects of matter and
radiation has the further benefit that it permits a simple check
of stability from a holographic point of view. Scalar fields on
asymptotically de Sitter spacetimes induce a conformal field
theory at infinity \cite{kn:stromingerdscft}\cite{kn:mazur}; this
is independent of the hypothetical existence of a complete
``dS/CFT correspondence". This CFT frequently has complex
conformal weights, which may possibly signal yet another
instability or other pathology. We show that this does not happen
in our model.

Finally, again guided by \cite{kn:OVV}, we shall discuss how the
singularities mentioned above can be resolved. This necessarily
involves a brief violation of the \emph{Null Ricci Condition
}[NRC]. The resulting overall structure emphasises the link with
string gas cosmology \cite{kn:brandvafa}\cite{kn:unstable}.

\addtocounter{section}{1}
\section*{\large{\textsf{2. Euclid to Lorentz According to Ooguri et al}}}
Ooguri et al work with a compactification of IIB string theory on
a Euclidean manifold of the form
$\bbr\,\times\,\m{S^1\,\times\,S^2\,\times\,CY}$, where CY is a
Calabi-Yau manifold, where S$^2$ has its usual metric of curvature
1/L$^2$, and where $\bbr\,\times\,\m{S}^1$ is a \emph{partial
compactification} of the hyperbolic surface of curvature
$-\,1$/L$^2$. Recall that one form of the metric on the simply
connected hyperbolic surface H$^2$ [topology $\bbr^2$] is
\begin{equation}\label{eq:PORCUPINE}
g(\m{H}^2)_{++}\;=\;\m{e^{(2\,\zeta/L)}\,d\varrho^2\;+\;d\zeta^2},
\end{equation}
where both $\varrho$ and $\zeta$ range in
($-\,\infty,\,+\,\infty$). Ooguri et al work with the space [of
topology $\bbr\,\times\,\m{S}^1$ and geometry H$^2/\bbz$] obtained
by compactifying $\varrho$ but \emph{not} $\zeta$.

Leaving aside the $\m{CY\,\times\,S^2}$ factor for the moment, we
have here a two-dimensional Euclidean space which Ooguri et al
interpret in \emph{two different ways}. In one, the circular
version of $\varrho$ is ``time", leading to a Witten index which
counts ground states of a certain black hole configuration. In the
other, the circular coordinate is ``space", and we have a
cosmological model, which Ooguri et al regard as evolving in
Euclidean ``time", $\zeta$, from an ``initial" S$^1$. As $\zeta$
increases, the size of the ``spatial" circle increases
exponentially, so, physically, what we should obtain in this way
is an \emph{accelerating cosmology}. Of course, in speaking of
``time" and ``space", we are really referring to some Lorentzian
version of this Euclidean space, so what Ooguri et al are
proposing is that one has a single Euclidean space with \emph{two
different Lorentzian versions}, one of which is a two-dimensional
accelerating cosmology. One of our objectives will be to clarify
how it is possible for a single Euclidean space to have more than
one Lorentzian version.

Meanwhile, however, let us note that Ooguri et al are really
proposing a kind of \emph{duality} between supersymmetric black
holes and accelerating cosmologies. If this can be made to work,
then we will have \emph{a new tool for studying accelerating
cosmologies} from a string-theoretic point of view, assuming that
such cosmologies can be obtained in string theory in some other
way \cite{kn:kklt}. The procedure would be to extend from two to
four dimensions by replacing H$^2/\bbz$ with its natural
higher-dimensional version, the partial compactification
H$^4/\bbz^3$ [obtained by compactifying the sections in the
foliation of H$^4$ by flat slices]; the most obvious replacement
for $\bbr\,\times\,\m{S^1\,\times\,S^2\,\times\,CY}$ would then be
$\bbr\,\times\,\m{T^3\,\times\,FR}$, where $\bbr\,\times\,$T$^3$
is the topology of H$^4/\bbz^3$, and where FR denotes a [singular]
Freund-Rubin space of the kind studied in
\cite{kn:bobby}\cite{kn:lambert}\cite{kn:valandro}.

This approach differs from that of Maldacena and Maoz
\cite{kn:maoz}, who have constructed string-motivated cosmological
models by introducing matter into locally \emph{anti-de Sitter}
spacetimes. These cosmologies can be made to accelerate
[temporarily], but only by introducing a quintessence field
\cite{kn:mcinnes}\footnote{See
\cite{kn:horowitz1}\cite{kn:answering}\cite{kn:reallyflat}\cite{kn:horowitz2}
for other studies of anti-de Sitter cosmology.}.

However, this is not what Ooguri et al have in mind. They hope to
obtain a direct interpretation of the H$^2/\bbz$ metric
\emph{itself} in terms of a cosmology which accelerates without
the aid of any quintessence field. Clearly this will only be
possible by means of some unusual mathematical approach, because
one normally associates hyperbolic metrics [constant
\emph{negative} curvature] with anti-de Sitter spacetime rather
than de Sitter spacetime. Furthermore, as Ooguri et al themselves
observe, it is far from clear how to interpret $\zeta$ in
(\ref{eq:PORCUPINE}) as Lorentzian time, because complexifying it
in the traditional way will complexify the metric. These two
problems will in fact prove to be two sides of the same coin.

The key idea here is that of \emph{multiple analytic
continuation}, as used in studies of ``bubble of nothing"
spacetimes. The idea [originally suggested in \cite{kn:bubble}] is
that it can be useful and physically meaningful to complexify
\emph{spacelike} as well as timelike dimensions of a Lorentzian
spacetime, provided that in the end one obtains a real metric with
physically reasonable signature. Thus, for example, it is
acceptable to complexify both a time and a space coordinate in
certain black hole and other solutions
\cite{kn:danny}\cite{kn:vijay}\cite{kn:jones} to obtain other
interesting spacetimes [``bubbles of nothing"]. Applied to the
Euclidean case, this clearly opens the way to obtaining two [or
more] different Lorentzian spacetimes from a single Euclidean
space, just as Ooguri et al require. To understand how this works
for asymptotically de Sitter and anti-de Sitter spacetimes, let us
first explain how it applies to dS and AdS themselves. After doing
so, we shall return to the specific case of the OVV spacetime at
the end of this section.

It is a basic fact that if one takes the standard metric on the
four-sphere of radius L,
\begin{eqnarray}\label{eq:D}
g(\mathrm{S^4})_{++++}\; =\;
\m{L}^2\,\Big\{\m{d\xi^2}\;+\;\m{cos^2(\xi)}\,[\mathrm{d\chi^2\;
+\; sin^2(\chi)}\{\mathrm{d}\theta^2 \;+\;
\mathrm{sin}^2(\theta)\,\mathrm{d}\phi^2\}]\Big\},
\end{eqnarray}
where all of the coordinates are angular, and continues
$\xi\,\rightarrow\,\m{iT/L}$, then the result is de Sitter
spacetime,
\begin{eqnarray}\label{eq:E}
g(\mathrm{dS_4})_{-\,+++}\; =\;-\,
\m{dT}^2\;+\;\m{L}^2\,\m{cosh^2(T/L)}\,[\mathrm{d\chi^2\; +\;
sin^2(\chi)}\{\mathrm{d}\theta^2 \;+\;
\mathrm{sin}^2(\theta)\,\mathrm{d}\phi^2\}],
\end{eqnarray}
with the indicated signature.

Now let us begin again, applying the ideas of ``multiple analytic
continuation" \cite{kn:bubble}. Let us complexify $\chi$ in
equation (\ref{eq:D}) instead of $\xi$, replacing
$\chi\,\rightarrow\,\m{\pm is/L}$, and for convenience relabelling
$\xi$ as u/L [without complexifying it]. This is a multiple
analytic continuation in Witten's \cite{kn:bubble} sense, that is,
it changes the sign of \emph{more than one} term in the metric. We
obtain
\begin{eqnarray}\label{eq:F}
g(\mathrm{AdS_4})_{+---}\; =\;
\m{du^2}\;-\;\m{cos^2(u/L)}\,[\mathrm{ds^2\; +\;\m{L^2}\,
sinh^2(s/L)}\{\mathrm{d}\theta^2 \;+\;
\mathrm{sin}^2(\theta)\,\mathrm{d}\phi^2\}].
\end{eqnarray}
\emph{But this is the anti-de Sitter metric}, in ($+\;-\;-\;-$)
signature, and expressed in terms of coordinates
\cite{kn:hawking}\cite{kn:gibbons}\cite{kn:orbifold} based on the
timelike geodesics which are perpendicular to the spatial
sections; the coordinate u is proper time along these geodesics.
These coordinates do not cover the entire spacetime, of course,
because these timelike geodesics intersect, being drawn together
by the attractive nature of gravity in anti-de Sitter spacetime
[which satisfies the Strong Energy Condition]. This is why these
coordinates give the false impression that there is no timelike
Killing vector in this geometry. In fact, these coordinates cover
the Cauchy development of a single spacelike slice. Thus we have
only continued S$^4$ to a small \emph{part} of AdS$_4$, so the
procedure is unsatisfactory in this sense; in particular, in the
opposite direction, this way of Euclideanizing AdS$_4$ would
obviously not be suitable for describing its asymptotic regions.
Nevertheless it is now clear that it is not correct to claim that
de Sitter spacetime is \emph{the only} Lorentzian continuation of
the four-sphere: one can even obtain a piece of anti-de Sitter
spacetime in this way. In fact, it is clear more generally that
Euclidean spaces will often have more than one Lorentzian version
if we accept both ($+\;-\;-\;-$) and ($-\;+\;+\;+$) signatures
--- as we must, since the difference
is a mere matter of convention\footnote{It is in fact possible to
dispute this statement in the case of non-orientable spacetimes
\cite{kn:carlip}, but we shall not consider this case here.}.

For many purposes, the more familiar $\xi\,\rightarrow\,\m{iT/L}$
complexification of S$^4$ is the most satisfactory way of thinking
about Euclidean de Sitter spacetime; one might well prefer it when
considering the \emph{earliest} history of the Universe. However,
there is one very crucial aspect of de Sitter spacetime that is
\emph{not} well described in this way: its asymptotic region. Just
as complexifying AdS$_4$ to S$^4$ is not a good way to study the
boundary physics of anti-de Sitter spacetime, so also we cannot
expect to understand the \emph{late-time}, near-boundary physics
of dS$_4$ by continuing it to S$^4$, which of course has no
boundary whatever. This argument suggests strongly that a full
understanding of asymptotically de Sitter spacetimes can only be
obtained by \emph{complementing} the standard Euclidean version of
de Sitter spacetime with a continuation to a Euclidean manifold
which, unlike S$^4$, can be regarded as the interior of a
manifold-with-boundary. One picture would be suitable for very
\emph{early} times, the other for very \emph{late} times.

The solution of this problem is to regard de Sitter spacetime as
the continuation not only of S$^4$, \emph{but also of the
hyperbolic manifold} H$^4$. This was proposed [in this context] in
\cite{kn:exploring}; explicit implementations were investigated in
\cite{kn:bala1} and \cite{kn:cadoni} [see also \cite{kn:bala2}];
it has been used in the very interesting theory of Lasenby and
Doran \cite{kn:lasenby1}\cite{kn:lasenby2}; it has been put on a
rigorous mathematical basis [though mainly in the case of Einstein
bulks, which are of limited cosmological interest] by Anderson
\cite{kn:anderson}; closely related ideas appear in the theory of
S-brane and ``bubble" spacetimes \cite{kn:jones}; it is relevant
to any theory which makes use of the fact that the de Sitter and
anti-de Sitter spacetimes are mutually locally conformal
\cite{kn:eva}; and, as we shall see, it plays a basic role in the
recent work of Ooguri, Vafa, and Verlinde \cite{kn:OVV}. Let us
explain why this suggestion is not as radical as it may seem.

First, one is [now] accustomed to think of dS$_4$ as a space of
constant \emph{positive} curvature, and hence it seems ``natural"
to associate it with S$^4$. However, this is merely due to the
[current] preference for signature ($-\;+\;+\;+$): if we use
($+\;-\;-\;-$) signature, then the curvature of dS$_4$ is
\emph{negative}. Similarly the curvature of the AdS$_4$ metric in
($+\;-\;-\;-$) signature, given in (\ref{eq:F}) above, is
\emph{positive}, as it should be since (\ref{eq:F}) was obtained
from the metric of the Euclidean four-sphere. Again, the overall
sign is a matter of convention, so there is no basis for the
assertion that de Sitter spacetime is ``naturally" associated with
S$^4$. Indeed, in some works [for example
\cite{kn:lasenby1}\cite{kn:lasenby2}] the opposite is assumed to
be natural.

A more serious objection is that AdS$_4$ can of course \emph{also}
be obtained by analytic continuation from H$^4$. Now AdS$_4$
differs from dS$_4$ not just in its local geometry but also in its
global structure. How can these different global structures arise
from different complexifications of a single Euclidean space?

First let us obtain the formal derivation. One global foliation of
H$^4$, shown in Figure 1, is in terms of cylinders with topology
$\bbr\times\m{S}^2$.
\begin{figure}[!h]
\centering
\includegraphics[width=0.7\textwidth]{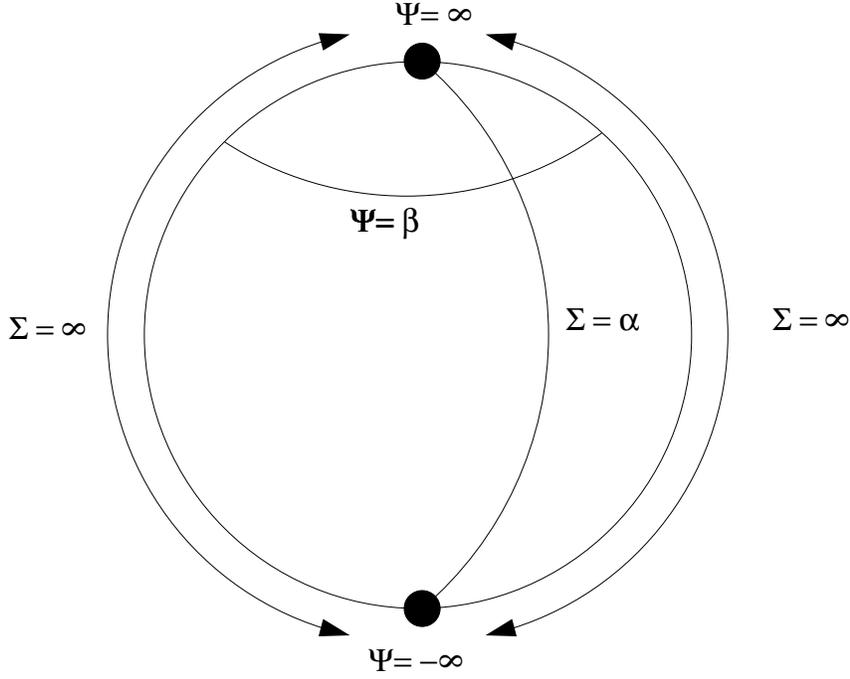}
\caption{Cylindrical foliation of H$^4$.}
\end{figure}
This works as follows. H$^4$ can be defined as a connected
component of the locus
\begin{equation}\label{eq:G}
\m{- \,A^2\;+\: B^2\; + \;x^2 \;+ \; y^2\; +\; z^2\; =\; -\,L^2},
\end{equation}
defined in a five-dimensional Minkowski space. It is clear that
all of the coordinates except A can range in
($-\,\infty,\;+\,\infty$), while A has to satisfy
$\m{A^2\;\geq\;L^2}$. Choosing the connected component on which A
is positive, we can pick coordinates
$\Psi$,$\Sigma$,$\theta$,$\phi$ such that
\begin{eqnarray} \label{eq:H}
\m{A} & = & \m{L\;cosh(\Psi)\;cosh(\Sigma)}                       \nonumber \\
\m{B} & = & \m{L\;sinh(\Psi)\;cosh(\Sigma)} \nonumber\\
\m{z} & = & \m{L\;sinh(\Sigma)\;cos(\theta)}           \nonumber \\
\m{y} & = & \m{L\;sinh(\Sigma)\;sin(\theta)\;cos(\phi)}  \nonumber \\
\m{x} & = & \m{L\;sinh(\Sigma)\;sin(\theta)\;sin(\phi)},
\end{eqnarray}
and these coordinates cover H$^4$ globally if we let $\Psi$ run
from $-\,\infty$ to $+\,\infty$ while $\Sigma$ runs from 0 to
$+\,\infty$. [Both $\theta$ and $\phi$ are suppressed in Figure
1.] The surfaces $\Psi$ = constant are copies of three-dimensional
hyperbolic space, \emph{all of the same curvature}, $-$1/L$^2$, as
can be seen by noting that the first two equations of (\ref{eq:H})
imply that $\m{-\,A^2\;+\;B^2}$ is independent of $\Psi$. These
slices intersect the conformal boundary at right angles. A typical
surface $\Psi$ = $\beta$ = constant is shown in Figure 1.

The surfaces $\Sigma$ = constant are topological cylinders. If we
think of H$^4$ as the interior of a four-dimensional ball, then
these cylinders are ``pinched" as they approach the boundary at
the points $\Psi\;=\;\pm\infty$. A typical ``pinched cylinder"
inside the boundary is shown as $\Sigma$ = $\alpha$ = constant in
Figure 1. It is clear that H$^4$ is completely foliated by these
cylinders, and the conformal boundary is a cylinder with two
additional points [corresponding to $\Psi\;=\;\pm\infty$] added:
with these additions, the boundary becomes the familiar
three-sphere. Notice that this structure is very similar to that
of Lorentzian anti-de Sitter spacetime. The metric with respect to
this foliation is
\begin{eqnarray}\label{eq:I}
g(\m{H^4})_{++++}\;=\; \m{L^2\,\Big\{cosh^2(\Sigma) \, d\Psi^2\;+
\;d\Sigma^2 \; +\; sinh^2(\Sigma)[d\theta^2 \; +\;
sin^2(\theta)d\phi^2]\Big\}}.
\end{eqnarray}

Now we shall consider a second, completely different, but also
entirely global foliation of H$^4$, shown in Figure 2.
\begin{figure}[!h]
\centering
\includegraphics[width=0.6\textwidth]{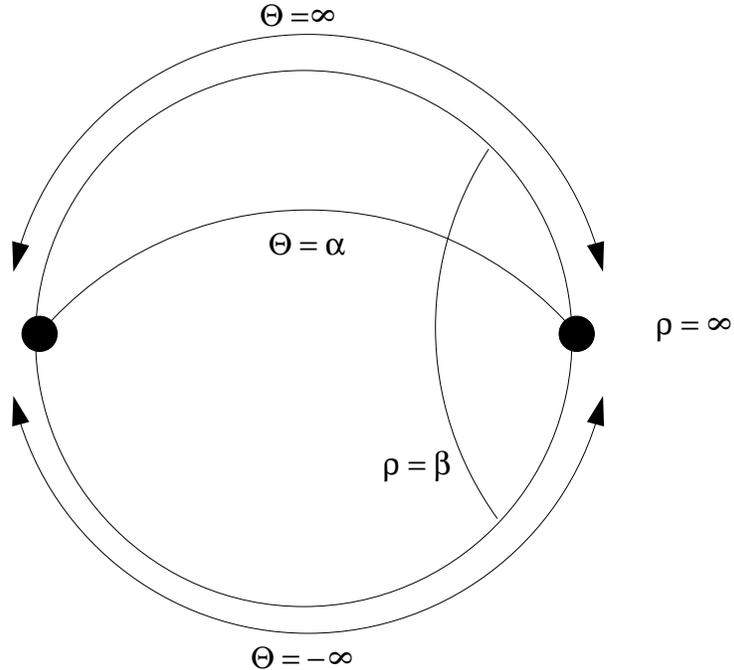}
\caption{H$^3$ foliation of H$^4$.}
\end{figure}
Choose coordinates $\Theta$,$\rho$,$\theta$,$\phi$, where $\Theta$
runs from $-\,\infty$ to $+\,\infty$ while $\rho$ runs from 0 to
$+\,\infty$, and set
\begin{eqnarray} \label{eq:J}
\m{A} & = & \m{L\;cosh(\Theta)\;cosh(\rho)}                       \nonumber \\
\m{B} & = & \m{L\;sinh(\Theta)} \nonumber\\
\m{z} & = & \m{L\;cosh(\Theta)\;sinh(\rho)\;cos(\theta)}           \nonumber \\
\m{y} & = & \m{L\;cosh(\Theta)\;sinh(\rho)\;sin(\theta)\;cos(\phi)}  \nonumber \\
\m{x} & = &
\m{L\;cosh(\Theta)\;sinh(\rho)\;sin(\theta)\;sin(\phi)}.
\end{eqnarray}
Because we are suppressing two angles, Figure 2 seems to resemble
Figure 1, but this is misleading [except in two dimensions, see
below]. Here the surfaces $\Theta$ = constant are, from the second
member of equations (\ref{eq:J}), just the submanifolds B =
constant; they are copies of the three-dimensional hyperbolic
space H$^3$, as can be seen at once from equation (\ref{eq:G}).
This foliation differs from the previous one in a crucial way,
however: whereas previously the slices all had the same curvature,
$-1$/L$^2$, as the ambient space, here the surface $\Theta$ =
$\alpha$ = constant can be written as
\begin{equation}\label{eq:JACKAL}
\m{- \,A^2\;+\; x^2 \;+ \; y^2\; +\; z^2\; =\;
-\,L^2\,cosh^2(\alpha)},
\end{equation}
so the magnitude of the curvature of a slice is reduced by a
factor of sech$^2$($\alpha$). The slices become flatter as they
are expanded towards the boundary. In this case, the copies of
H$^3$ are all ``pinched together" as we move towards \emph{their}
boundaries, that is, as $\rho\,\rightarrow\,\infty$. A typical
H$^3$ slice, $\Theta$ = $\alpha$ = constant is shown in Figure 2.

Notice that the slices themselves do not intersect: only their
conformal completions do so. At any point on a given copy of
H$^3$, one can send a geodesic [shown in Figure 2] of the form
$\theta$ = $\phi$ = constant, $\rho$ = constant = $\beta$, towards
infinity, and this will uniquely define two points on the
boundary, one each at $\Theta = \pm\infty$. In this sense, one can
say that conformal infinity is ``disconnected": the usual
three-sphere is divided into two hemispheres corresponding to the
forward or backward ``evolution" of any H$^3$ slice along
$\Theta$. Of course, topologically the boundary is connected,
since the two hemispheres join along the common conformal boundary
of all of the slices. Nevertheless, it will be useful to remember
that the boundary does fall into two disconnected pieces if the
boundary of the slices is deleted.

It is clear that this foliation is also global, though it is in
general totally different to the one shown in Figure 1. The metric
with respect to this foliation is
\begin{eqnarray}\label{eq:K}
g(\mathrm{H}^4)_{++++}\; =\;\m{L^2}\,\Big\{
\m{d\Theta^2}\;+\;\m{cosh^2(\Theta)}[\mathrm{d\rho^2}\; +\;
\mathrm{\,sinh^2(\rho)}\{\mathrm{d}\theta^2 \;+\;
\mathrm{sin}^2(\theta)\,\mathrm{d}\phi^2\}]\Big\}.
\end{eqnarray}
[See for example \cite{kn:randall} for a different application of
this foliation of H$^4$.]

Now the metrics in (\ref{eq:I}) and (\ref{eq:K}) are one and the
same; only the foliations are different. It has been stressed by
Buchel and Tseytlin \cite{kn:tseytlin}\cite{kn:buchel}, however,
that it is not the case that two distinct foliations of a given
Euclidean space must be regarded as fully equivalent in the
quantum theory. This is true even in the case where different
foliations correspond, on the Lorentzian side, to different
foliations of the spacetime by spacelike slices with distinct
intrinsic geometries. For this will correspond to distinct time
evolutions with physically distinct quantum Hamiltonians. [On the
Lorentzian side, different spacetime foliations correspond to
different groups of observers with non-intersecting worldlines, so
one should think of this in terms of observer dependence rather
than violation of diffeomorphism invariance.] One should think of
``Euclideanization" as an assignment to a given Lorentzian
manifold of a Euclidean manifold \emph{with a fixed foliation.} In
our case the distinction between the two foliations pictured is
even more stark, since, after complexification, the leaves of one
foliation become spacelike while those of the other are themselves
Lorentzian. Thus we should not in general regard Figures 1 and 2
as depictions of the same physical system, even though the
underlying Euclidean manifold is the same in both cases.
\begin{figure}[!h]
\centering
\includegraphics[width=0.5\textwidth]{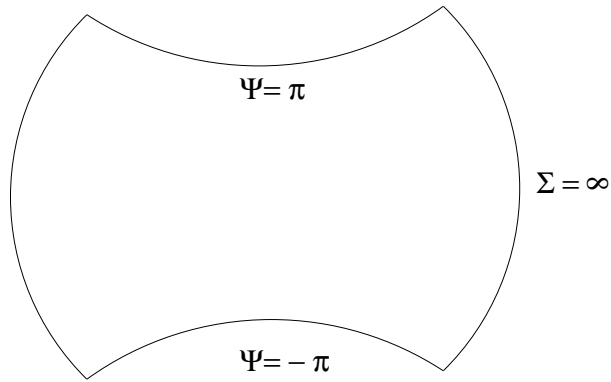}
\caption{Cylindrical foliation of H$^4$, with compactification of
one axis.}
\end{figure}

Furthermore, different foliations of a given Euclidean space can
be appropriate to different ways of taking quotients by discrete
isometry groups. For example, it is quite natural to compactify
the coordinate $\Psi$ in equation (\ref{eq:I}), since
$\Psi\,\rightarrow\,\Psi\,+\,2\pi$ is clearly an isometry. If we
do this, then Figure 1 has to be re-drawn; parts of the diagram
have to be deleted, as shown in Figure 3, and the top and bottom
of the diagram have to be identified. The boundary changes
topology from S$^3$ to S$^1\,\times\,\m{S}^2$. This is just the
``thermal AdS" construction which is of such importance in the
AdS/CFT correspondence \cite{kn:wittenads}. By contrast, in
(\ref{eq:K}) neither $\Theta$ nor $\rho$ can be compactified in
this way, at least not in four dimensions [see below]. Hence this
foliation is appropriate to the case where we do \emph{not} take
the quotient. However, the \emph{local} geometry of the two, now
distinct, Euclidean spaces remains identical.

Let us now proceed to the complexifications. For the sake of
clarity we shall adopt the convention that a non-periodic
\emph{time} coordinate becomes periodic, and vice versa, upon
complexification; note that this is compatible both with our
discussion above and with the usual procedure when continuing, for
example, the Schwarzschild metric. There is however no reason to
insist on this rule when dealing with \emph{spacelike}
coordinates. In that case we shall try to preserve the periodicity
or non-periodicity of the coordinate, unless the geometry is such
that it seems natural to do otherwise.

In view of our earlier discussion, we should not expect
(\ref{eq:I}) and (\ref{eq:K}) necessarily to lead to the same
Lorentzian spacetime upon complexification. Indeed, if we map
$\Psi\,\rightarrow\,\m{iU/L}$ and re-label $\Sigma$ [without
complexifying it] as S/L, then we obtain from (\ref{eq:I})
\begin{eqnarray}\label{eq:L}
g(\m{AdS_4)_{-\,+++}\;=\; -\,cosh^2(S/L) \, dU^2\;+ \;dS^2 \; +\;
L^2\,sinh^2(S/L)[d\theta^2 \; +\; sin^2(\theta)d\phi^2]},
\end{eqnarray}
and this is precisely \cite{kn:hawking} the \emph{globally valid}
AdS$_4$ metric, in the indicated signature\footnote{Unless we take
the universal cover, the topology of AdS$_4$ is
S$^1\,\times\,\bbr^3$, so the time coordinate U is, in the first
instance, a circular coordinate; see
\cite{kn:hawking}\cite{kn:gibbons}\cite{kn:orbifold}. According to
our rule, we pass to the universal cover in the Lorentzian case if
and only if $\Psi$ is periodic.}. But if we take (\ref{eq:K}) and
perform a multiple analytic continuation [again, in Witten's sense
\cite{kn:bubble} of changing the sign of more than one term in the
metric] by mapping $\rho\,\rightarrow\,\pm \m{i\chi}$ while
re-labelling $\Theta$ as T/L, we obtain, since sinh($\pm$
i$\,\chi$) = $\pm \m{i\;sin(\chi)}$,
\begin{eqnarray}\label{eq:M}
g(\m{dS_4)_{+---}\; =\; dT^2\;-\;L^2\,cosh^2(T/L)\,[d\chi^2\; +\;
sin^2(\chi)\{d\theta^2 \;+\; sin^2(\theta)\,d\phi^2\}]},
\end{eqnarray}
which is of course the \emph{global} form of the de Sitter metric,
but in a signature which makes its curvature \emph{negative}. Thus
de Sitter and anti-de Sitter spacetimes are seen to have a common
origin in the same Euclidean space, which has however been
foliated in two ways that, in four dimensions, are very
different\footnote{If we compactify $\Psi$ then we should say that
they come from Euclidean spaces with the same \emph{metric} but
with different \emph{topologies}.}.

Notice that, just as the cylindrical structure of the foliation in
Figure 1 is very much like the cylindrical structure of Lorentzian
anti-de Sitter spacetime, so also the structure of Figure 2 is
very closely related to that of Lorentzian de Sitter spacetime, in
the following sense: at each point of a spacelike section of
Lorentzian de Sitter spacetime, one can send out a timelike
geodesic either to the future or the past, and so map the point to
a unique point on either future or past conformal infinity. We saw
that a precisely analogous statement could be made regarding
Figure 2. The difference is that the conformal infinity of
Lorentzian de Sitter spacetime is truly disconnected, whereas the
Euclidean ``future" and ``past" in Figure 2 bend around and touch
along the equator, so that the full conformal infinity is one
connected copy of S$^3$. If one were to delete this equator then
the formal similarity would be still closer.

We are not of course claiming that these simple observations
immediately allow a full understanding of de Sitter spacetime in
terms of [say] the Euclidean AdS/CFT correspondence. Our original
motivation \cite{kn:exploring} for relating de Sitter spacetime to
hyperbolic space was the observation that, from the point of view
of symmetries, H$^4$ is much more similar to dS$_4$ than to
AdS$_4$. Indeed, H$^4$ and dS$_4$ have exactly the same isometry
group, the orthogonal group O(1,4). Furthermore, the conformal
boundary of dS$_4$ consists of two copies of S$^3$, which of
course is also the conformal boundary of H$^4$. These facts led
us, in \cite{kn:exploring}, to propose that one or both of the
boundary components of dS$_4$ should be topologically identified
with the boundary of H$^4$, thus enabling the symmetries of one
side to act directly on the other. A concrete proposal for
relating the physics on one side to that on the other was then
proposed in \cite{kn:bala1} [see also\cite{kn:bala2}]. The
relationship is non-local, and the transformation has a
non-trivial kernel, so that information is lost as one tries to
transfer de Sitter data to H$^4$. Thus one cannot establish an
\emph{exact equivalence} of the AdS/CFT kind here. Nevertheless it
is certainly reasonable to expect that gross features of the
physics on one side, such as the stability of an entire spacetime
and its matter content, are reflected on the other.

All of our discussions thus far have been relevant to
four-dimensional spacetimes. In other dimensions the situation is
similar, though as we are about to see the case of two spacetime
dimensions is rather subtle. However, the reader may wish to note
the following point. Euclideanization of curved metrics was of
course developed in connection with Euclidean Quantum Gravity
\cite{kn:HG}, but the method has taken on a life of its own,
particularly in connection with the AdS/CFT correspondence.
Nevertheless, if one is interested in the original application,
then it is important that complexification of a coordinate should
also complexify the volume form. In equation (\ref{eq:K}), for
example, the volume form is
\begin{eqnarray}\label{eq:MONSTER}
\m{dV}(g(\mathrm{H}^4)_{++++})\; =\;\m{L^4\,
cosh^3(\Theta)\,sinh^2(\rho)\,sin(\theta)\,d\Theta\,d\rho\,d\theta\,d\phi},
\end{eqnarray}
and one sees at once that complexifying $\rho$ [to obtain equation
(\ref{eq:M})] does indeed complexify the volume form. However,
this only works if the number of spacetime dimensions is
\emph{even}. Thus we shall confine ourselves to even spacetime
dimensions henceforth. [Depending on the dimension, one may have
to choose the sign of the imaginary factor in the complexification
so that the volume form ``rotates" in the correct direction. This
is the reason for the $\pm$ sign in the complexification of
$\rho$, above.]

In two dimensions, the situation we have been describing is
particularly interesting. In that case, Figures 1 and 2 can be
interpreted literally, in the sense that there are no angles to be
suppressed. Then a simple reflection, $\Psi\,\rightarrow\,\rho$,
$\Sigma\,\rightarrow\,\Theta$ shows that the two foliations are
identical, and indeed it is clear that the two ways of writing the
Euclidean metric are the same:
\begin{equation}\label{eq:N}
g\m{(H^2)_{++}\;=\;L^2\,[cosh^2(\Sigma)\,d\Psi^2\;+\;d\Sigma^2]}\;=\;\m{L^2}\,[
\m{d\Theta^2}\;+\;\m{cosh^2(\Theta)}\,\mathrm{d\rho^2}];
\end{equation}
here, though not in higher dimensions, we can take $\rho$ to be
periodic, like $\Psi$. [We are therefore dealing with the quotient
H$^2/\bbz$, and the correct Euclidean picture is then the one in
Figure 3, or Figure 3 reflected about a diagonal if one prefers
the other foliation.]

Complexifying as usual, we obtain
\begin{equation}\label{eq:O}
g(\m{AdS_2)_{-\,+}} \;=\; \m{-\,cosh^2(S/L) \, dU^2\;+ \;dS^2},
\end{equation}
\begin{equation}\label{eq:OCELOT}
g\m{(dS_2)_{+\,-}\;=\; dT^2\;-\;L^2\,cosh^2(T/L)\,d\chi^2}.
\end{equation}
Now it is immediately clear that the two-dimensional case has a
special property: \emph{it is not possible to use the metric to
decide which coordinate should be regarded as time}. For, locally,
these two metrics are identical. Ultimately this is due to the
fact that the anti-de Sitter group in n+1 dimensions, O(2,n), is
isomorphic to the de Sitter group O(n+1,1), when n = 1. The local
geometry does not tell us whether equation (\ref{eq:O}) is the
AdS$_2$ metric with U as time, or (\ref{eq:OCELOT}) is the dS$_2$
metric with T as time [and a suitable choice of signature]. This
corresponds precisely to the fact that the two Euclidean
foliations are identical. However, our rule regarding the
periodicity of time under complexification helps us: if we take
$\Psi$ to be periodic, then, if U is time, it should \emph{not} be
periodic: this gives us the standard topology $\bbr^2$ for
two-dimensional anti-de Sitter spacetime. On the other hand, if
$\chi$ is spacelike, then it should continue to be periodic, and
this gives us the standard $\bbr\,\times\,\m{S}^1$ topology for
two-dimensional de Sitter spacetime. Thus the global structure
helps to remove the ambiguity.

In this approach, the distinction between AdS$_2$ and dS$_2$
cannot be seen in the Euclidean theory: both spacetimes arise from
the Euclidean space pictured in Figure 3. \emph{The distinction
arises from the choice as to which dimension is to be interpreted
as time in the Lorentzian continuation}. Notice that neither
interpretation involves complexifying the metric; nor does either
lead to a periodic Lorentzian time: the periodicity is pushed off
to Euclidean time in the AdS case, and to Lorentzian space in the
dS case. This indicates that our rule as to when a complexified
time coordinate should be periodic does have a firm physical
basis.

In this work we shall be primarily concerned not with the de
Sitter and anti-de Sitter spacetimes, but rather with spacetimes
which are asymptotic to these fundamental examples. The above
discussion was however necessary, because we must take care to
perform analytic continuations in a way which is compatible with
the continuation of the asymptotic spacetime. The basic point is
that we now know how to continue an asymptotically hyperbolic
Euclidean space to a cosmological spacetime. \emph{This is
precisely what is done in the work of Ooguri, Vafa, and Verlinde}
\cite{kn:OVV}, to which we now turn.

As we discussed earlier, the two-dimensional hyperbolic space can
be foliated in yet a third way, by flat sections: the metric was
given as (\ref{eq:PORCUPINE}). After compactifying one direction,
as in \cite{kn:OVV}, the manifold becomes H$^2/\bbz$, and the
metric becomes
\begin{equation}\label{eq:P}
g(\m{H}^2/\bbz)_{++}\;=\;\m{K^2\,e^{(2\,\zeta/L)}\,d\vartheta^2\;+\;d\zeta^2};
\end{equation}
here, $\vartheta$ as an angular coordinate. This partial
compactification introduces a new length scale, denoted K, which
is independent of the curvature scale L. This additional length
scale is of basic importance.

As Ooguri et al point out, however, it is difficult to see how
complexification can work in equation (\ref{eq:P}) if we proceed
in the usual way, since complexifying $\zeta$ leads to a complex
metric. Our earlier discussion reveals the correct procedure: if
we think of $\vartheta$ as Euclidean ``time" and complexify it as
$\vartheta\,\rightarrow\,\m{i\tau}$/K [re-labelling $\zeta$ as z],
we obtain a \emph{non-periodic} time coordinate $\tau$ on [part
of] ordinary AdS$_2$:
\begin{equation}\label{eq:Q}
g(\m{AdS^*_2)_{-\,+}\;=\; -\,e^{2z/L} \, d\tau^2\;+ \;dz^2};
\end{equation}
while if we think of K as a measure of the size of Euclidean
``space", then the continuation K $\,\rightarrow\,$ $\pm$iK, with
a re-labelling of $\zeta$ as t and of $\vartheta$ as $\theta_1$,
gives us a local de Sitter spacetime:
\begin{equation}\label{eq:R}
g(\m{dS}^*_2/\bbz)_{+\,-}\;=\;\m{
dt^2\;-\;K^2\,e^{2t/L}\,d\theta_1^2}.
\end{equation}
Here the asterisks remind us that the foliations of
\emph{Lorentzian} AdS and dS by flat sections do not cover the
entire spacetimes. Furthermore, these flat sections are not
compact in the original spacetime, so if [in accordance with our
rule] the coordinate $\theta_1$ is periodic, then we are dealing
with dS$^*_2/\bbz$ here, as indicated. This space happens to have
the same $\bbr\,\times\,\m{S}^1$ topology as the standard
two-dimensional de Sitter spacetime, but it is constructed quite
differently: it is obtained\footnote{See the next Section for a
discussion of this construction in the four-dimensional case.} by
cutting away part of dS$_2$, leaving a geodesically incomplete
spacetime, and then patching this up by means of a topological
identification. \emph{The fact that one returns to the original
topology in this way is merely a quirk of two-dimensional
geometry.} This observation will be crucial below; meanwhile, we
stress that, while they may have the same metric, AdS$_2^*$ and
$\m{dS}^*_2/\bbz$ are globally very different. This is how one
distinguishes the two different Lorentzian spacetimes which Ooguri
et al wish to obtain from a single Euclidean space.

We have obtained a consistent picture of the proposal of Ooguri et
al, \emph{one which avoids the problem of complex metrics}. The
analysis also instructs us how to work towards more realistic,
four-dimensional cosmological models. We should begin with the
four-dimensional version of the Euclidean space H$^2/\bbz$, namely
H$^4/\bbz^3$, where, as always, H$^{\m{n}}$ denotes the standard
simply connected space of constant negative curvature
$-\,1$/L$^2$; here, we take the H$^4$ foliation by simply
connected flat surfaces, as described for example in \cite{kn:oz},
and compactify these surfaces as we did in the two-dimensional
case. The simplest choice is a cubic torus with all circumferences
equal to position-dependent multiples of 2$\pi$K; other shapes and
topologies are possible and interesting but will not be considered
here. With this construction we obtain the manifold
$\bbr\,\times\,\m{T}^3$ with a metric which is just the obvious
extension of $g(\m{H}^2/\bbz)_{++}$ [equation (\ref{eq:P}) above].
This is a space of constant negative curvature $-$1/L$^2$; it is a
partial compactification of H$^4$. Therefore one can study in it
all of the usual Euclidean AdS physics, such as the Seiberg-Witten
\cite{kn:seiberg} brane pair creation process.

We are interested here in the cosmological Lorentzian version of
the ``Euclidean OVV space", H$^4/\bbz^3$. In accordance with the
above discussion, we should therefore complexify K in the same way
as in the two-dimensional case; since this will change the
signature from ($+\;+\;+\;+$) to ($+\;-\;-\;-$) [because K is the
radius of all three circles], this can be regarded as another
example of Witten's \cite{kn:bubble} multiple analytic
continuation. In fact, we \emph{must} proceed in this way in order
to ensure compatibility with the ($+\;+\;+\;+$) to ($+\;-\;-\;-$)
complexification of the basic H$^4$ with which we began. We
finally have the four-dimensional version of the OVV metric,
\begin{equation}\label{eq:T}
g_{\m{OVV}}(\m{K,\,L)_{+---} \;=\; dt^2\;
-\;K^2\,e^{(2\,t/L)}\,[d\theta_1^2 \;+\; d\theta_2^2 \;+\;
d\theta_3^2]}.
\end{equation}
This is a Lorentzian metric with the indicated signature: thus it
is still a metric of constant \emph{negative} curvature, but it is
\emph{locally} indistinguishable from the ($+\;-\;-\;-$) version
of dS$_4$ with its foliation by flat surfaces. As discussed
earlier, complexifying K $\rightarrow\,\pm$ iK will complexify the
volume form [which is proportional to K$^{\m{n}}$ in (n+1)
dimensions] provided that the number of spacetime dimensions is
even; thus the path integral can be made to converge in the
four-dimensional case if the sign of the imaginary factor is
suitably chosen.

We have argued that, if the Ooguri et al proposals lead to a
reasonable four-dimensional cosmology at all, then the basic form
of this cosmology is a certain version of de Sitter spacetime,
with the metric (\ref{eq:T}), and with a distinctive
\emph{topology}. Because of its fundamental importance, we need to
study the properties of this spacetime in detail.

\addtocounter{section}{1}
\section*{\large{\textsf{3. Structure and Stability of the OVV Spacetime}}}
In this section we shall explore the very interesting global
structure of the four-dimensional OVV spacetime, and we shall show
that it is stable against the Seiberg-Witten pair production
process.

First we consider the details of the construction. Consider the
version of ($+\;-\;-\;-$) de Sitter spacetime with flat spatial
sections. Physically, this is precisely the set of all events in
dS$_4$ to which a signal can be sent by an inertial observer. This
is in fact an open submanifold of $\bbr\,\times\,\m{S}^3$: the
Penrose diagram \cite{kn:aguirre} is obtained by deleting from the
usual dS$_4$ diagram the lower triangular half, \emph{including
the diagonal and its endpoints}. This is shown as the triangle OAC
in Figure 4; the fact that the endpoints of the diagonal have been
deleted is indicated by the small circles. Notice that timelike
geodesics perpendicular to the flat spatial sections are
represented by straight lines converging on the point O.
\begin{figure}[!h]
\centering
\includegraphics[width=0.6\textwidth]{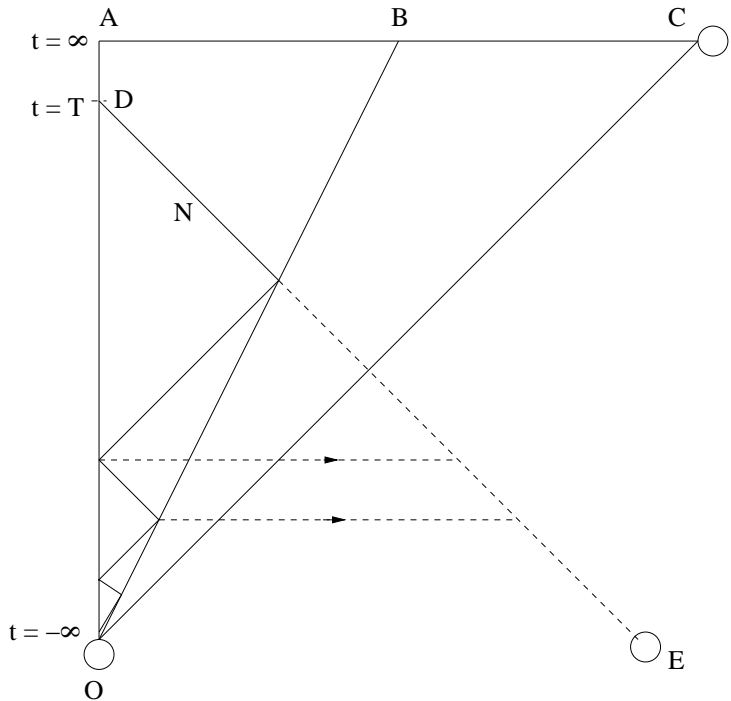}
\caption{Penrose diagram of the OVV spacetime.}
\end{figure}
It is clear from the diagram that \emph{this spacetime is timelike
and null geodesically incomplete}, but we can \emph{partly} remedy
this by compactifying the spatial sections. This works as follows.

In the Penrose diagram, a typical point represents a two-sphere.
We can however inscribe a cube into each two-sphere, with the
cubes in different two-spheres having parallel faces, and then
think of the points in the diagram in terms of the inscribed
cubes. Since cubes do not of course have spherical symmetry, we
will have to interpret the Penrose diagram as representing a
specific direction. Now in the Penrose diagram of the half of de
Sitter spacetime which can be foliated by copies of flat $\bbr^3$,
take the cube of side length 2$\pi$K at time t = 0 and perform the
usual identifications to generate a three-torus. The effect is to
remove all of the diagram to the right of the line marked OB,
which is the timelike geodesic corresponding to this cube. Thus
the Penrose diagram corresponding to the OVV metric is just the
triangle OAB shown, where however the point O is \emph{not}
included.

Clearly the line OB does not otherwise intersect the diagonal line
OC where the original spacetime was incomplete, so the only place
where the new spacetime can be incomplete is the point O. This
point is at t = $-\,\infty$, so it is at an infinite proper time
to the past along any timelike curve; thus this spacetime is
timelike geodesically complete. However, consider the null curve N
shown, which begins at t = T and extends back to t = $-\,\infty$.
Because of the topological identifications, it appears to
``bounce" back and forth between the origin and OB: actually it
wraps around the torus, which of course shrinks as we move back in
time. It will in fact wrap around the torus infinitely many times.
To compute the amount of affine parameter expended as it moves
towards O, proceed as follows: first, write equation (\ref{eq:T})
as
\begin{equation}\label{eq:U}
g_{\m{OVV}}(\m{K,\,L)_{+---}
\;=\;e^{(2\,t/L)}\,\{d\eta^2\;-\;K^2[d\theta_1^2 \;+\; d\theta_2^2
\;+\; d\theta_3^2]\}},
\end{equation}
where $\eta$ is conformal time; thus dt = $\m{e^{(t/L)}\,d\eta}$.
As the conformally related metric in the braces is flat, a null
geodesic satisfies $\m{d\eta/d\widetilde\lambda}$ = c$_1$ =
constant, where $\widetilde\lambda$ is an affine parameter with
respect to this flat metric. The usual relation [\cite{kn:wald},
page 446] between the affine parameters of null geodesics of
conformally related metrics gives us
$\m{d\widetilde\lambda/d\lambda \;=\;c_2\,e^{-\,(2\,t/L)}}$, where
c$_2$ is another constant. Thus we have
\begin{equation}\label{eq:V}
\m{{{dt}\over{d\lambda}}\;=\;{{dt}\over{d\eta}}\,{{d\eta}\over{d\widetilde\lambda}}\,
{{d\widetilde\lambda}\over{d\lambda}}
\;=\;e^{(t/L)}\,c_1\,c_2\,e^{-\,(2\,t/L)}\;=\;c\,e^{-\,(t/L)}}.
\end{equation}
Now it should be clear from the construction of the line DE in
Figure 4 that the ``corners" in the null geodesic N are not
relevant to the computation of the affine parameter A(N) along N.
The answer is simply given by
\begin{equation}\label{eq:W}
\m{A(N)\;=\;c\,\int_{-\,\infty}^T\,e^{(t/L)}dt},
\end{equation}
which is obviously \emph{finite}. Thus we have explicitly
constructed a null geodesic which is inextensible and yet expends
finite total affine parameter as it is extended back towards the
point O. We conclude that the OVV spacetime is \emph{timelike but
not null geodesically complete}.

One might think that this incompleteness is a special property of
the specific metric $g_{\m{OVV}}(\m{K,\,L)_{+---}}$, but this is
\emph{not} the case. In fact, a theorem\footnote{\emph{Not} the
one used in \cite{kn:unstable}.} due to Andersson and Galloway
\cite{kn:andergall}\cite{kn:gall} implies that null incompleteness
to the past is almost unavoidable here. The theorem may be stated
as follows; we refer the reader to the Appendix for details of the
terminology and a brief commentary on this remarkable result.

\bigskip
\noindent THEOREM [Andersson-Galloway]: \emph{Let} M$_{\m{n+1}}$,
n $\leq$ 7, \emph{be a globally hyperbolic (n+1)-dimensional
spacetime with a regular future spacelike conformal boundary}
$\Gamma^+$. \emph{Suppose that the Null Ricci Condition is
satisfied and that} $\Gamma^+$ \emph{is compact and orientable.
If} M$_{\m{n+1}}$ \emph{is past null geodesically complete, then
the first homology group of} $\Gamma^+$, H$_1(\Gamma^+,\,\bbz)$,
\emph{is pure torsion}.
\bigskip

Here the \emph{Null Ricci Condition} is the requirement that, for
all null vectors k$^{\mu}$, the Ricci tensor should satisfy
\begin{equation}\label{eq:AA}
\mathrm{R}_{\mu\nu}\,\mathrm{k}^\mu\,\mathrm{k}^\nu\;\geq\;0.
\end{equation}

The OVV spacetime has a regular future conformal boundary,
pictured as the line AB in Figure 4; this boundary is compact and
orientable, since it is just the torus T$^3$. The Null Ricci
Condition is satisfied, since the spacetime is an Einstein space.
But the first homology group of T$^3$ is certainly \emph{not} pure
torsion [that is, not every element is of finite order]: it is
isomorphic to $\bbz\,\oplus\,\bbz\,\oplus\,\bbz$. Thus the
spacetime \emph{had} to be null geodesically incomplete.

The surprising and beautiful feature of the Andersson-Galloway
theorem, however, is that this same conclusion holds \emph{even if
we distort the geometry}, as long as the Null Ricci Condition
continues to hold. That is, if we introduce any kind of matter
such that the NRC continues to hold --- if the Einstein equations
are valid, then this just means that the Null Energy Condition
holds --- then the resulting cosmology will necessarily be null
incomplete to the past. We can interpret the theorem in this way
because, physically, the introduction of matter into the OVV
spacetime will not violate its other conditions. In other words,
the Andersson-Galloway theorem is a \emph{singularity theorem}: as
with the classical singularity theorems, the conclusion does not
depend on assumptions about local isotropy or homogeneity or on
the precise structure of the Friedmann equations. The behaviour it
predicts will occur no matter what happens to the spatial sections
in the early Universe.

Obviously the null incompleteness of the OVV spacetime itself is
rather harmless, since clearly there is no ``singularity" at O. It
is true that, in principle, an observer in this spacetime can
receive a message from the ``negative infinite" past; but since
one can receive messages in ordinary AdS$_4$ from spatial
infinity, perhaps this no longer seems very shocking. More
seriously, however, one should note that this spacetime contains
nothing but dark energy. If we were to introduce some ordinary
matter or radiation satisfying the NRC, then we would certainly
expect a real singularity to develop. The situation here is
analogous to that of pure AdS$_4$, with metric given in equation
(\ref{eq:L}). This space satisfies the \emph{Strong} Energy
Condition [SEC], and yet it is not singular. This does not
contradict the Hawking-Penrose theorem, however, but only because
AdS$_4$ does not satisfy the \emph{generic condition}
[\cite{kn:hawking}, page 101], a technical condition on the
curvature tensor. In the same way, the null geodesic
incompleteness of the OVV spacetime will turn into a genuine
curvature singularity if we introduce any kind of matter such that
the Null Ricci Condition continues to hold, because the spacetime
will become generic in this sense. This expectation is confirmed
by the explicit examples we shall consider later.

We stress that this result is \emph{very much stronger} than the
classical singularity theorems, which require the SEC. For it is
easy to violate the SEC with well-behaved matter such as
quintessence, but it is extremely difficult to do so in the case
of the Null Energy Condition, a fact which has been discussed from
many different points of view in the recent literature --- see for
example
\cite{kn:entropy}\cite{kn:carroll}\cite{kn:hsu}\cite{kn:unstable}\cite{kn:latta}.
This is particularly true if any NEC violation persists beyond the
earliest history of the Universe. We must nevertheless accept that
any concrete cosmological model based on the OVV spacetime will be
singular if we only consider non-exotic matter fields and if the
Einstein equations hold exactly. This point will be discussed in
more detail below.

Let us return to the OVV spacetime itself. Recall that the Penrose
diagram is just the triangle OAB in Figure 4. Evidently, like de
Sitter spacetime, the OVV spacetime has a cosmological horizon,
which would extend diagonally down from point A in the Figure.
Clearly, however, the observer at the origin can receive signals
from any point in space\footnote{See
\cite{kn:smash}\cite{kn:leblond} for early discussions of such
phenomena.}; indeed, as we saw, he can see rays of light which
have circumnavigated his world arbitrarily many times. In reality,
of course, he can see back no farther than the time of decoupling,
but, even in this case, we can adjust K [which is inversely
related to the slope of the line OB in Figure 4] so that he will
\emph{eventually} be able to see an entire spatial section even if
--- apparently like ourselves \cite{kn:cornish}\cite{kn:menzies} --- he cannot do
so at present. At that time he will be able to deduce the fate of
all physical systems in his world, despite the fact that they are
destined ultimately to pass beyond the horizon. Thus the horizon
has a rather different, less questionable status in this
spacetime. See \cite{kn:bousso}\cite{kn:unstable} for discussions
of related issues.

Finally, let us consider the stability of this spacetime from the
Seiberg-Witten \cite{kn:seiberg} point of view. The fact that the
Euclidean OVV space is locally indistinguishable from Euclidean
AdS$_4$, which is of course stable in this sense, should not make
us complacent. For branes, being extended objects, are sensitive
to certain \emph{global} geometric features: their actions
typically involve non-local quantities, such as area and volume,
the growth of which can be strongly affected by the global
structure of the ambient manifold. One might be concerned that
these global phenomena might affect area and volume differently.
This does in fact happen in the OVV case, though fortunately not
to the extent that any instability is induced.

Consider a BPS (D $-$ 1)-brane together with an appropriate
antisymmetric tensor field in a Euclidean asymptotically
AdS$_{\mathrm{D} + 1}$ background. The brane action consists
\cite{kn:seiberg} of two terms: a positive one contributed by the
brane tension, but also a \emph{negative} one induced by the
coupling to the antisymmetric tensor field. As the first term is
proportional to the area of the brane, while the second is
proportional to the volume enclosed by it, we have
\begin{equation}\label{eq:X}
\mathrm{S} \;=\;
\mathrm{T}(\mathrm{A}\;-\;{{\mathrm{D}}\over{\mathrm{L}}}\,\mathrm{V}),
\end{equation}
where T is the tension, A is the area, V the volume enclosed, and
L is the background asymptotic AdS radius. In the case at hand,
[the Euclidean version of] equation (\ref{eq:T}) gives us
\begin{equation}\label{eq:XYLOPHONE}
\mathrm{S_{OVV}(t)} \;=\;
8\pi^3\mathrm{T\,K^3}\,\Big\{\mathrm{e^{3\,t/L}}\;-\;{{3}\over{\mathrm{L}}}\,
\int_{-\,\infty}^{\m{t}}\mathrm{e^{3\,\tau/L}\,d\tau}\Big\},
\end{equation}
which of course is \emph{precisely zero}. In reality, as we
discussed above, this spacetime will become singular if we
introduce non-exotic matter into it, so we should really begin the
integration at some finite value, say t = $-\,\sigma$; then the
action is the positive constant
$8\pi^3\mathrm{T\,K^3\,e^{-\,3\sigma/L}}$. The introduction of
matter will however modify the geometry, and so the action will in
general either increase or decrease away from this value as t is
taken to infinity. If it becomes negative, then we have a
non-perturbative instability of the system. Explicit examples of
this have been given in \cite{kn:maoz} and \cite{kn:unstable}.

Clearly the OVV spacetime is not unstable in this sense; however,
it would not be difficult to render it unstable, since even a
small constant negative contribution to the slope of the action
could eventually lead to negative values for the action itself.
Evidently we need to consider the back-reaction on the OVV
spacetime geometry arising from the presence of matter. This is
the subject of the next section.

\addtocounter{section}{1}
\section*{\large{\textsf{4. Structure of OVV Cosmologies Containing Non-Exotic
Matter}}}

In this section we shall consider the consequences of introducing
``non-exotic" matter into the OVV spacetime studied in the
previous section. Here ``non-exotic" means that the matter shall
be such that the corresponding stress-energy tensor satisfies the
Null Energy Condition [NEC]. \emph{If} we assume that the Einstein
equations hold, this is equivalent to assuming that the spacetime
geometry satisfies the Null Ricci Condition\footnote{See
\cite{kn:OBSERVATION} for discussions of the observational aspects
of NEC violation, and \cite{kn:NECVIOLATION} for the theoretical
aspects.}. However, the reader is reminded that in more general
contexts, such as brane world models
\cite{kn:varun1}\cite{kn:varun2}\cite{kn:coley} or Gauss-Bonnet
models \cite{kn:sasaki}, it is possible for the NRC to be violated
even if all true matter fields satisfy the NEC. This is known as
``effective" violation of the NEC; see \cite{kn:nojiri} for a
general discussion of this point. This will be important later,
but, throughout this section, we shall assume that the Einstein
equations do hold.

Our specific concern in this section is to understand the
classical structure of the spacetimes obtained by deforming the
OVV spacetime, using non-exotic matter. Questions of stability are
postponed to the next section.

We proceed as follows. We take the OVV spacetime, with a small
cosmological constant determined by a length scale L, and
introduce into it a scalar field $\varphi$ with a potential
\begin{equation}\label{eq:Y}
\m{V(\varphi,\;\epsilon)\;=\;-\,{{3}\over{8\pi
L^2}}\,[1\;-\;{{1}\over{6}}\,\epsilon]\,sinh^2(\sqrt{2\,\pi\,\epsilon}\,\varphi)};
\end{equation}
here $\epsilon$ is a positive constant. We have placed the minus
sign prominently so as to remind the reader that, in the
($+\;-\;-\;-$) signature being used here, the sign of a potential
is the opposite of the familiar one. The kinetic term does not
change sign, however. Bear in mind that OVV spacetime is
\emph{locally} the same as de Sitter spacetime, which in this
signature has a negative cosmological constant $-\,3/(8\pi$L$^2$).

While we shall also be interested in the specific behaviour of a
small scalar field excitation on an OVV background, our primary
concern at this point is with the following claim: we assert that,
as far as the spacetime geometry is concerned, the field $\varphi$
\emph{exactly mimics} the effects on the OVV spacetime of a matter
field with \emph{constant} equation-of-state parameter related to
$\epsilon$ by
\begin{equation}\label{eq:Z}
\m{w_{\varphi}\;=\;{{1}\over{3}}\,\epsilon\;-\;1}.
\end{equation}
Thus for example if we insert non-relativistic matter [zero
pressure] into the OVV spacetime and allow it to act [via the
Einstein equations], this will have the same effect as introducing
$\varphi$ with $\epsilon$ = 3, while $\varphi$ with the value
$\epsilon$ = 4 mimics the effects of radiation; the value
$\epsilon$ = 1 arises if we are interested in the back-reaction
induced by domain walls on the OVV geometry, and so on. We stress
that we are \emph{not} primarily interested in using this field to
violate the Strong Energy Condition; that is, $\varphi$ is not
[necessarily] a quintessence field. In the cases of principal
interest to us, the acceleration is due to the negative
contribution made by the OVV cosmological constant to the total
pressure. Also note that we are \emph{not} claiming that $\varphi$
is a fundamental scalar field: we use it as a convenient way of
representing various kinds of matter to be inserted into the OVV
spacetime. If we take the kinetic term to be the standard one,
then, as is well known, $\varphi$ automatically satisfies the Null
Energy Condition, so the condition that the matter should be
``non-exotic" holds for any value of the parameters.

We now proceed to justify these claims. We shall consider
Friedmann-like cosmological models with metrics of the form
\begin{equation}\label{eq:ALPHA}
g \;=\; \m{\m{dt}^2\; -\; \m{K^2\;a(t)}^2[d\theta_1^2 \;+\;
d\theta_2^2 \;+\; d\theta_3^2]};
\end{equation}
this generalizes the OVV metric in an obvious way. Adding the
energy density of the $\varphi$ field to that of the initial OVV
space, we have a Friedmann equation of the form
\begin{equation}\label{eq:BETA}
\m{({{\dot{a}\over{a}}})^2 \;=\;
{{8\pi}\over{3}}\;[\;{{1}\over{2}}\;\dot{\varphi}^2
\;-\;\textup{V}(\varphi,\;\epsilon) \;+\;{{3}\over{8\pi L^2}}]}.
\end{equation}
The equation for $\varphi$ itself is
\begin{equation}\label{eq:GAMMA}
\m{\ddot{\varphi}\;+\;3\;{{\dot{a}}\over{a}}\;\dot{\varphi}\;-
\;{{dV(\varphi,\;\epsilon)}\over{d\varphi}}\;=\;0.}
\end{equation}
Surprisingly, these equations have very simple solutions: by
muddling about in the manner of \cite{kn:mcinnes}, one finds that
[with natural initial conditions] $\varphi$ is given by
\begin{equation}\label{eq:DELTA}
\m{\varphi\;=\;{{1}\over{\sqrt{\pi\epsilon/2}}}\,tanh^{-\,1}(e^{-\,\epsilon\,t/2L})},
\end{equation}
and the metric is\footnote{The reader who wishes to undertake the
task of verifying these solutions will find the following simple
fact helpful: if A and B are quantities related by tanh(A) =
e$^{-\m{B}}$, then cosh(2A) = coth(B).}
\begin{equation}\label{eq:EPSILON}
g\m{_s(\epsilon,\,K,\,L)_{+---} \;=\; dt^2\; -\;
K^2\;sinh^{(4/\epsilon)}({{\epsilon\,t}\over{2L}})\,[d\theta_1^2
\;+\; d\theta_2^2 \;+\; d\theta_3^2]};
\end{equation}
here the s refers to the sinh function. From these results one can
compute the energy density and pressure of the $\varphi$ field
alone:
\begin{equation}\label{eq:ZETA}
\m{\rho_{\varphi}\;=\;{{3}\over{8\pi
L^2}}\,cosech^2({{\epsilon\,t}\over{2L}})},
\end{equation}
\begin{equation}\label{eq:ETA}
\m{p_{\varphi}\;=\;{{3}\over{8\pi
L^2}}\,[{{1}\over{3}}\,\epsilon\;-\;1]\,cosech^2({{\epsilon\,t}\over{2L}})},
\end{equation}
from which equation (\ref{eq:Z}) above is immediate.

If $\epsilon$ = 3, we should have the local metric for a spacetime
containing non-relativistic matter and a de Sitter cosmological
constant, and indeed $g\m{_s(\epsilon,\,K,\,L)_{+---}}$ reduces in
this case
--- purely locally, of course --- to the classical Heckmann metric
[see \cite{kn:overduin} for a recent discussion]. In the general
case it agrees [again locally] with the results reported in
\cite{kn:eroshenko}, where it is obtained by postulating a linear
equation of state [without giving a matter model]. For large t we
have
\begin{equation}\label{eq:THETA}
g\m{_s(\epsilon,\,2^{\,2/\epsilon}\,K,\,L)_{+---} \;\approx\;
dt^2\; -\;K^2\,e^{2\,t/L}[d\theta_1^2 \;+\; d\theta_2^2 \;+\;
d\theta_3^2]},
\end{equation}
which is the OVV metric given in equation (\ref{eq:T}); notice
that $\epsilon$ effectively drops out. Thus our metric is
``asymptotically OVV", for all $\epsilon$.

The matter content of this spacetime does not behave as simply as
one might expect. For while it is true that both components, the
cosmological constant and the $\varphi$ field, separately have
constant equation-of-state parameters, \emph{their combination
does not}: if we denote the total energy density by $\rho$ and the
total pressure by p, then we have a total equation-of-state
parameter w given by
\begin{equation}\label{eq:IOTA}
\m{w\;=\;p/\rho\;=\;-\,1\;+\;{{\epsilon}\over{3}}\,
sech^2({{\epsilon\,t}\over{2L}})}.
\end{equation}
Thus w decreases from $-\,1\;+\;(\epsilon/3)$ in the early
universe to its asymptotic OVV value $-\,1$. The Strong Energy
Condition is violated if w $<$ $-\,1$/3, so the SEC holds in the
early universe provided that $-\,1\;+\;(\epsilon/3)\, >\, -\,1/3$,
which just means that $\epsilon$ should exceed 2. In this case
there is a transition from deceleration to acceleration, as is
observed in our Universe \cite{kn:riess}. This is of course the
case of most interest.

Clearly the field $\varphi$ diverges as we trace it back towards t
= 0, and so do its energy density and pressure. It follows from
the Einstein equations that this spacetime has a genuine
[curvature] singularity there: for example, the scalar curvature
is given by
\begin{equation}\label{eq:IZOLA}
\m{R}(g\m{_s(\epsilon,\,K,\,L)_{+---}) \;=\;
-\,{{12}\over{L^2}}\;+\;{{3}\over{L^2}}\,(\epsilon\;-\;4)\,cosech^2({{\epsilon
t}\over{2L}})};
\end{equation}
this tends to $-$12/L$^2$ as t tends to infinity, the correct
asymptotic de Sitter value in this signature, but it clearly
diverges as t tends to zero [except in the $\epsilon$ = 4 case,
which corresponds to radiation and hence to a traceless
stress-energy tensor which does not contribute to the scalar
curvature; but this case is still singular, as one sees by
examining other curvature invariants]. \emph{This cannot be
understood in terms of the classical singularity theorems},
because some of the metrics in this family violate the
\emph{Strong} Energy Condition even at early times, and yet remain
singular: for example, the metric
\begin{equation}\label{eq:IOTAIOTA}
g\m{_s(2,\,K,\,L)_{+---} \;=\; dt^2\; -\;
K^2\;sinh^2({{t}\over{L}})\,[d\theta_1^2 \;+\; d\theta_2^2 \;+\;
d\theta_3^2]}
\end{equation}
violates the SEC at \emph{all} times t $>$ 0, and yet it is
singular. The same statement is true for all members of the family
with $\epsilon$ less than 2. This is in contrast to spacetimes
with the $\bbr\,\times\,\m{S^3}$ topology of [the global, simply
connected version of] de Sitter spacetime, where of course the
classical singularity theorems can be evaded precisely because de
Sitter spacetime itself violates the SEC. That is, SEC violation
is enough to remove the singularity in that case, \emph{but not
here}.

Instead, we have to use the Andersson-Galloway theorem
\cite{kn:andergall}\cite{kn:gall}, which applies to these
spacetimes in exactly the same way as we applied it to the OVV
spacetime in the preceding section. That is, the failure of null
geodesic completeness is an inevitable consequence of the
\emph{topology} of these spacetimes [the first homology group of
the spatial sections is not pure torsion] combined with the fact
that $\varphi$ automatically satisfies the \emph{Null} Energy
Condition [which is equivalent to the Null Ricci Condition here
since we are assuming the Einstein equations]. But, as we foresaw,
the failure of null geodesic completeness here is more serious
than in the pure OVV case: the spacetime curvature is no longer
constant, the geometry is generic [in the technical sense
discussed earlier], and the result is a curvature singularity,
which enforces timelike as well as null geodesic incompleteness.
The Andersson-Galloway theorem implies that this singularity can
only be avoided in one way: by violating the NRC. This important
conclusion will be discussed in more detail later.

\begin{figure}[!h]
\centering
\includegraphics[width=0.3\textwidth]{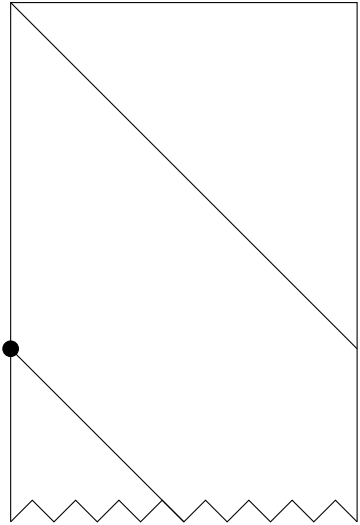}
\caption{Penrose diagram of the $g\m{_s(\epsilon,\,K,\,L)_{+---}}$
spacetime, $\epsilon
> 2$}
\end{figure}

The complete structure of these spacetimes is summarized, as
usual, in their Penrose diagrams, which are of two types according
to the value of $\epsilon$. The Penrose diagram for the
$\epsilon\,>\,2$ case is shown in Figure 5. Future conformal
infinity is spacelike, and there is a Big Bang singularity which
is also spacelike. The height/width ratio of the diagram is
determined by the parameter ratio L/K; we have chosen a value such
that the observer shown cannot \emph{yet} detect the toral
structure of his universe, though he will detect it later. The
reader may find it helpful to compare Figure 5 with the diagrams
in \cite{kn:bousso}\cite{kn:chiba}.

If $\epsilon\,\leq\,2$, the Penrose diagram is triangular, like
the triangle OAB in Figure 4, with the difference that the point O
becomes genuinely singular. These diagrams and their precise
shapes will be discussed in more detail elsewhere
\cite{kn:challenge}. Meanwhile, we remind the reader that we are
really most interested in values of $\epsilon$ between 3 [pure
non-relativistic matter] and 4 [pure radiation].

We now turn to questions of stability.

\addtocounter{section}{1}
\section*{\large{\textsf{5. Stability of OVV Cosmologies Containing Non-Exotic
Matter}}}

In this section we shall be concerned with the stability of these
spacetimes, specifically from a holographic/string-theoretic point
of view.

We have treated the $\varphi$ field merely as a device for
representing the effects of various kinds of matter and radiation
on the OVV spacetime. However, let us ask what happens if we take
it more seriously, as a genuine scalar field excitation
propagating on the OVV background [without affecting the spacetime
geometry].

The field theory of scalars on de Sitter-like backgrounds is a
vast subject; see for example
\cite{kn:gazeau}\cite{kn:tolley}\cite{kn:dolgov} for relevant
work. A new aspect of the theory was revealed, however, by studies
of de Sitter spacetime from a holographic point of view. It was
soon realised \cite{kn:stromingerdscft}\cite{kn:mazur} that a
scalar field propagating in the de Sitter bulk induces a conformal
field theory on the conformal boundary. Whether this implies the
existence of a complete \emph{equivalence} of the AdS/CFT kind is
questionable \cite{kn:bala1}\cite{kn:bala2}, but we shall not rely
on the existence of a complete ``dS/CFT correspondence" here. We
merely wish to ask what kind of CFT is induced by our field
$\varphi$.

In detail, the limit of a scalar field amplitude [for a scalar of
mass m$_{\varphi}$] in de Sitter spacetime defines a CFT two-point
function at de Sitter conformal infinity. The conformal weights
corresponding to the boundary operator defined by $\varphi$ are
given, in four spacetime dimensions, by
\begin{equation}\label{eq:XI}
\m{h_{\pm}\;=\;{{1}\over{2}}\,[3\;\pm\;\sqrt{9\;-\;4\,L^2\,m_{\varphi}^2}}].
\end{equation}
One sees immediately that the weights will be \emph{complex}, with
unwelcome physical consequences, unless m$_{\varphi}$ satisfies
m$_{\varphi}^2 \,\leq\,9/(4$L$^2$). This ``Strominger bound" is
analogous
\cite{kn:stromingerdscft}\cite{kn:exploring}\cite{kn:cadoni} to
the well-known Breitenlohner-Freedman bound \cite{kn:freed} on the
masses of scalar fields in anti-de Sitter spacetime.

Now at late times, the metric $g\m{_s(\epsilon,\,K,\,L)_{+---}}$
[equation (\ref{eq:EPSILON})] is locally indistinguishable from
that of de Sitter spacetime, and, furthermore, $\varphi$ is very
small [equation (\ref{eq:DELTA})]; hence we see, from
(\ref{eq:Y}), that $\varphi$ can be regarded as a scalar field of
squared mass
\begin{equation}\label{eq:NU}
\m{m_{\varphi}^2\;=\;{{3}\over{2L^2}}\,\epsilon\,[1\;-\;{{1}\over{6}}\,\epsilon]}
\end{equation}
propagating on a local de Sitter background. Substituting this
into equation (\ref{eq:XI}), we find that \emph{the Strominger
bound is automatically satisfied for all values of} $\epsilon$.
Both weights are always real, and they are given simply by
\begin{eqnarray} \label{eq:OMICRON}
\m{h}_{+} & = & \epsilon/2   \nonumber \\
\m{h}_{-} & = & 3\;-\;(\epsilon/2);
\end{eqnarray}
here we have assumed that $\epsilon\,\geq\,3$ so that
$\m{h}_+\,\geq\,\m{h}_-$; of course the definitions should be
reversed if $\epsilon\,<\,3$. Notice that by varying $\epsilon$
one can obtain all values of the mass allowed by the Strominger
bound. Recall that the action V($\varphi,\;\epsilon$) was
constructed simply with a view to obtaining equation (\ref{eq:Z});
so it is remarkable indeed that the construction is also precisely
what is required to ensure that the Strominger bound is satisfied.

We conclude that the field $\varphi$ always induces a well-behaved
CFT at infinity; our matter models seem to be acceptable from a
holographic point of view.

We complete this discussion with the following curious
observation. One sees from (\ref{eq:OMICRON}) that the Strominger
bound is saturated when $\epsilon$ = 3. Now because de Sitter
spacetime and Euclidean hyperbolic space have the same isometry
group, the relevant representation theory has been extensively
developed \cite{kn:vilenkin}. The scalar representations fall into
three families, the principal, complementary, and discrete series.
The principal representations are those which, under contraction
of the de Sitter group to the Poincar\'e group, correspond to the
familiar flat space representations \cite{kn:gazeau}. They are of
two kinds, which in fact are classified precisely by the weights
given above. The first kind is the case of complex weights: these
of course violate the Strominger bound. The only other kind is
precisely the case where the bound is saturated.

The complementary series consists of representations where the
Strominger bound is satisfied but not saturated\footnote{The
relevant Hermitian form in the complementary case
[\cite{kn:vilenkin}, page 518] involves a gamma function which is
ill-defined if the Strominger bound is saturated, so $\epsilon$ =
3 certainly does not belong to this series.}, but these
representations have no flat-spacetime analogue. [The discrete
series corresponds to the special, massless case
\cite{kn:tolley}\cite{kn:dolgov}.] We conclude that $\epsilon$ = 3
is the \emph{only} value which is both physically acceptable and
at the same time corresponds to a well-defined flat space
representation. It is interesting that $\epsilon$ = 3 is singled
out in this way, for this is physically perhaps the most important
case: as we saw, it corresponds to zero-pressure, non-relativistic
matter superimposed on an initial OVV spacetime.

Next we turn to our principal concern, the question of stability
from the point of view of the Seiberg-Witten pair production
process.

The Euclidean versions of these spacetimes are obtained by
complexifying K as usual, so we have
\begin{equation}\label{eq:KAPPA}
g\m{_s(\epsilon,\,K,\,L)_{++++} \;=\; dt^2\; +\;
K^2\;sinh^{(4/\epsilon)}({{\epsilon\,t}\over{2L}})\,[d\theta_1^2
\;+\; d\theta_2^2 \;+\; d\theta_3^2]}.
\end{equation}
This metric has, of course, the same asymptotic structure as the
Euclidean OVV metric
$g_{\m{OVV}}(\m{2^{-(2/\epsilon)}K,\,L)_{++++}}$; it is an example
of an asymptotically hyperbolic Riemannian metric. We can think of
it as a deformation of the OVV metric. This is of course alarming,
since, as was discussed in the Introduction, there are known
examples of such deformations [equation (\ref{eq:A})] which are
strongly unstable against pair production of BPS branes
\cite{kn:unstable}. We can investigate this issue by examining the
action given in equation (\ref{eq:X}): we have
\begin{equation}\label{eq:LAMBDA}
\m{S}(g\m{_s(\epsilon,\,K,\,L)_{++++}\,;\,t)\;=\;
8\pi^3\,TK^3[sinh^{(6/\epsilon)}({{\epsilon\,
t}\over{2L}})\;-\;{{3}\over{L}}\int_0^t
sinh^{(6/\epsilon)}({{\epsilon \,\tau}\over{2L}})\,d\tau ]}.
\end{equation}
In the case of non-relativistic matter [$\epsilon$ = 3] this can
be evaluated, the result being
\begin{equation}\label{eq:LLAMA}
\m{S}(g\m{_s(3,\,K,\,L)_{++++}\,;\,t)\;=\;
4\pi^3\,TK^3[{{3t}\over{L}}\;+\;e^{(3t/L)}\;-\;1]}.
\end{equation}
It is easy to see that this is \emph{never negative}, and so there
is no danger of Seiberg-Witten \cite{kn:seiberg} instability here.
In fact, this function increases away from its initial value,
zero. Thus the deformation is innocuous here, as it was \emph{not}
in the case of the metric $g\m{_c(2,\,2K,\,L)_{++++}}$ [equation
(\ref{eq:A})].

In the general case, a simple calculation shows that the
derivative of the brane action is given by
\begin{equation}\label{eq:MU}
\m{{{d}\over{dt}}\,S}(g\m{_s(\epsilon,\,K,\,L)_{++++}\,;\,t)\;=\;{{24\pi^3\,TK^3}\over{L}}
\,sinh^{(6/\epsilon)}({{\epsilon\, t}\over{2L}})[coth({{\epsilon\,
t}\over{2L}})\;-\;1]}.
\end{equation}
This is obviously positive. Since
$\m{S}(g\m{_s(\epsilon,\,K,\,L)_{++++}\,;\,0)}$ is zero, we see
that the action will never be negative, and so there is no
Seiberg-Witten instability here for any value of $\epsilon$. All
forms of matter that satisfy the Null Energy Condition help to
stabilize the spacetime.

Recall that the brane action was constant in the case of pure OVV
spacetime. What we have shown is that the introduction of matter
satisfying the NRC actually makes the OVV space ``more stable", in
the sense that it converts this constant to an increasing
function. By contrast, one can show \cite{kn:unstable} that
NRC-violating geometry causes it to \emph{decrease}. However, that
is not conclusive, since a decreasing function can of course still
be everywhere positive. This too was investigated in
\cite{kn:unstable}, and the conclusion is that a NRC-violating
geometry \emph{can} be stable, but only for certain values of the
parameters. We shall discuss this in more detail in the next
section.

To summarize, then, we have arrived at a somewhat paradoxical
conclusion. The OVV spacetime [pictured as the triangle OAB in
Figure 4] is well-behaved, but it is on the brink of two very
different catastrophes. First, it is ``nearly" singular in the
sense that it is null geodesically incomplete, but only in a
rather harmless way. Second, it is ``nearly" unstable in the sense
that a slight perturbation could set off the Seiberg-Witten
``stringy instability". We have found that making the OVV
spacetime more realistic --- by introducing conventional matter
and radiation into it --- definitely staves off instability, yet
at the same time causes the spacetime to become genuinely
singular. Furthermore, this singularity is of a particularly
persistent kind: no amount of SEC violation by a scalar with a
conventional kinetic term, or distortion of the geometry, can
remove it.

\addtocounter{section}{1}
\section*{\large{\textsf{6. Avoiding the Singularity }}}
We have stressed that the singularity in the metric
$g\m{_s(\epsilon,\,K,\,L)_{+---}}$ [equation (\ref{eq:EPSILON})]
is not a mere consequence of the assumed symmetries; nor however
can it be understood in terms of the classical singularity
theorems based on the Strong Energy Condition. In fact, the only
way\footnote{See the Appendix.} to remove this singularity is to
violate the Null Ricci Condition
\cite{kn:andergall}\cite{kn:gall}, or the Null Energy Condition if
we assume the validity of the Einstein equations.

A fundamental matter field which genuinely violates the NEC is
very hard to handle, as has been discussed for example in
\cite{kn:entropy}\cite{kn:carroll}\cite{kn:hsu}\cite{kn:unstable}\cite{kn:latta}.
It is true that, to avoid a singularity, we need only a brief
period of NEC violation, but it would clearly be better to avoid
the various unpleasant physical properties of NEC-violating
matter.

It is here that the distinction between the NRC and the NEC is
useful. The former is of course a purely geometric condition [see
(\ref{eq:AA}) above], while the latter refers only to items such
as pressure and energy densities. The two are linked by the
gravitational field equation. They are equivalent if the Einstein
equations hold exactly, but of course this is a highly
questionable assumption in the early Universe. In braneworld
models \cite{kn:varun1}\cite{kn:varun2}\cite{kn:coley} there are
explicit corrections to the Einstein equation which allow the NRC
to be violated \emph{while every matter field satisfies the NEC},
so that one says that the NEC is ``effectively" violated
\cite{kn:nojiri}\cite{kn:unstable}. Similarly, the NEC and the NRC
can be usefully different in certain Gauss-Bonnet theories
\cite{kn:sasaki}.

If we arrange to violate the NEC only effectively, then we can
hope to circumvent the Andersson-Galloway theorem without using
dangerously unstable forms of matter. However, there is a problem
here: the Seiberg-Witten instability is determined purely by
geometric data [rates of growth of volume and area] so it can
still be present \emph{even if the NEC is only violated
effectively}. This fact is the most serious obstacle to removing
the singularity at t = 0 in $g\m{_s(\epsilon,\,K,\,L)_{+---}}$,
and the same comment holds for any spacetime with the OVV
topology.

The question, then, is whether we can find an OVV-like spacetime
in which all matter fields preserve the NEC, in which the metric
violates the NRC, but which does not suffer from uncontrolled
Seiberg-Witten pair-production. Ideally one should do this by
deriving such a metric from a specific theory, such as in the
brane-world models mentioned above
\cite{kn:varun1}\cite{kn:varun2}\cite{kn:coley}; but for clarity
we shall just work with a simple family of metrics that violate
the NRC, and try to determine whether there are parameter values
allowing us to avoid any instability. A family of such metrics,
defined on a space with OVV topology $\bbr\,\times\,\m{T}^3$, was
discussed in \cite{kn:unstable}, and will now be described.

Consider the metrics given by
\begin{equation}\label{eq:PI}
g\m{_c(\gamma,\,K,\,L)_{+---} \;=\; dt^2\; -\;
K^2\;cosh^{(4/\gamma)}({{\gamma\,t}\over{2L}})\,[d\theta_1^2 \;+\;
d\theta_2^2 \;+\; d\theta_3^2]},
\end{equation}
where $\gamma$ is a positive constant and the c subscript refers
to the cosh function. Here K is simply the minimal radius of the
spatial torus. The similarity to
$g\m{_s(\epsilon,\,K,\,L)_{+---}}$ is obvious; in particular, this
spacetime is ``asymptotically OVV". The Penrose diagram is again
rectangular, as in Figure 5, with height determined by L and width
determined by K; the only difference is that the bottom of the
diagram is spacelike \emph{but not singular}. Since these
spacetimes satisfy all of the other conditions of the
Andersson-Galloway theorem and have a future conformal infinity
with toral topology, we see that they must violate the Null Ricci
Condition. In fact, these metrics were introduced in
\cite{kn:smash} precisely in order to demonstrate explicitly that
``phantom" cosmologies \cite{kn:NECVIOLATION} need not be
singular. They are obtained in the same way as
$g\m{_s(\epsilon,\,K,\,L)_{+---}}$, that is, by introducing
``matter" with a \emph{constant} equation-of-state parameter into
a de Sitter-like background; the difference is that this ``matter"
has negative energy density. These spacetimes \emph{always} [for
all $\gamma$ and at all times] violate the NRC. Thus they are a
kind of NRC-violating, non-singular ``dual" to the
$g\m{_s(\epsilon,\,K,\,L)_{+---}}$ spacetimes.

Consider the Seiberg-Witten brane action
$\m{S}(g\m{_c(\gamma,\,K,\,L)_{++++}\,;\,t)}$ for the Euclidean
version of this metric, $g\m{_c(\gamma,\,K,\,L)_{++++}}$, obtained
as usual by complexifying K. Clearly, since the area of a
cross-section is never zero,
$\m{S}(g\m{_c(\gamma,\,K,\,L)_{++++}\,;\,0)}$ is positive; with
tension T, the value is $8\pi^3\mathrm{T\,K^3}$. However, again
oppositely to the case in which the NRC is satisfied, the action
immediately decreases, and in fact it always decreases for all
values of t $>$ 0. The question as to whether it remains positive
as t $\rightarrow \,\infty$ is therefore settled by computing the
limiting value of the action. Unlike the action itself, this limit
can be evaluated explicitly. It turns out \cite{kn:unstable} that
it is actually negatively divergent for all values of
$\gamma\;\leq\;3$. [This is why the metric discussed in the
introduction, $g\m{_c(2,\,K,\,L)_{++++}}$, suffers from
Seiberg-Witten instability.] For values of $\gamma\;>\;3$ one can
show that
\begin{equation}\label{eq:VARPI}
\lim_{\m{t} \rightarrow
\infty}{\m{S}(g\m{_c(\gamma,\,K,\,L)_{++++}\,;\,t)}}\;=\;
2^{(3\;-\;{{6}\over{\gamma}})}\,\pi^3\,\m{K^3\,T}\,
{{(\gamma\;-\;6)(3\;+\;\gamma)}\over{\gamma(\gamma\;-\;3)}}.
\end{equation}
This is indeed negative for values of $\gamma$ strictly between 3
and 6. The spacetime will be unstable for these values, even if
the NEC violation is merely effective. But for values of $\gamma$
greater than or equal to 6, the limit is either positive or zero,
and therefore the action is strictly positive. Thus for example
the spacetime with OVV topology and metric
\begin{equation}\label{eq:RHO}
g\m{_c(6,\,K,\,L)_{+---} \;=\; dt^2\; -\;
K^2\;cosh^{(2/3)}({{3\,t}\over{L}})\,[d\theta_1^2 \;+\;
d\theta_2^2 \;+\; d\theta_3^2]}
\end{equation}
violates the NRC and yet is entirely stable against brane
pair-production: in fact the brane action in this case is simply
\begin{equation}\label{eq:VARRHO}
\m{S}(g\m{_c(6,\,K,\,L)_{++++}\,;\,t)\;=\;8\pi^3K^3T\,e^{(-\,3t/L)}}.
\end{equation}
Thus we see that we can indeed violate the NRC without
destabilizing the spacetime, and that this does remove the
singularity.

Since we wish to use this construction to remove the singularity,
it is natural to assume that it is relevant only to the very
earliest, pre-inflationary stages. Therefore we wish to choose the
parameters such that the spacetime naturally evolves to an
inflationary metric with the correct inflationary curvature
radius, L$_{\m{INF}}$. The OVV metric itself naturally has this
[local] inflationary structure if we choose L = L$_{\m{INF}}$ ---
that is, the OVV metric is [locally] just the one that is usually
called the inflationary metric. Furthermore, we know that the
metrics we have been considering do evolve to the OVV metric, so
we automatically have all the ingredients we need. We must however
determine how to choose the parameter $\gamma$ so that the
NRC-violating phase is brief.

If we take the unit timelike vector associated with proper time in
equation (\ref{eq:PI}) then we can, with the help of an arbitrary
unit spacelike vector perpendicular to it, construct a null vector
k$^{\mu}$; then it can be shown \cite{kn:unstable} that the
corresponding Ricci tensor satisfies
\begin{equation}\label{eq:SIGMA}
\mathrm{R}_{\mu\nu}\,\mathrm{k}^\mu\,\mathrm{k}^\nu\;=\;\m{{{-\,\gamma}\over{
L^2}}\,sech^2({{\gamma t}\over{2L}})}.
\end{equation}
The function on the right side therefore measures the extent of
NRC violation. For fixed L and t, this expression tends to zero as
$\gamma$ becomes large, yet its integral from 0 to $\infty$ is
$-$2/L, independent of $\gamma$; thus the effect of taking
$\gamma$ to be large is to focus the NRC violating effect close to
t = 0. Since ``large" values of $\gamma$ --- values greater than
or equal to 6 --- are precisely what we need for stability, we can
summarize this discussion as follows: NRC violation does not
destabilize the spacetime \emph{provided that it takes place over
a period of time which is sufficiently brief}. Since the magnitude
of R$_{\mu\nu}\,\mathrm{k}^\mu\,\mathrm{k}^\nu$ falls to half of
its maximal value in a time
\begin{equation}\label{eq:GADZOOKS}
\m{t_{1/2}\;=\;{{2\,L}\over{\gamma}}\,cosh^{-\,1}(\sqrt{2})},
\end{equation}
we see that, to be specific, instability will be averted if we can
arrange for the NRC-violating effect to decay so rapidly that
R$_{\mu\nu}\,\mathrm{k}^\mu\,\mathrm{k}^\nu$ falls to half its
maximal magnitude in a time no longer than
$\m{(L/3)\,cosh^{-\,1}(\sqrt{2})}$. As we have stressed, a brief
period of NRC violation is exactly what we want in an inflationary
picture, where L = L$_{\m{INF}}$; in fact, we might prefer it to
be even shorter than $\m{(L_{\m{INF}}/3)\,cosh^{-\,1}(\sqrt{2})}$,
that is, we might like $\gamma$ to be even larger than 6. However,
the value 6 is singled out as the smallest one which is physically
acceptable. The precise value of $\gamma$ has to be fixed by
quantum gravitational considerations.

Let us assemble the pieces of the puzzle. The basic idea is that,
in some model [such as a braneworld model] in which the Einstein
equations are corrected at very early times, the NEC and the NRC
initially fail to coincide : all of the matter fields obey the NEC
at all times, but the NRC is violated at these early times. For a
spacetime with the OVV topology, the result is a non-singular
metric which resembles $g\m{_c(\gamma,\,K,\,L_{INF})_{+---}}$.
There will be no instabilities if $\gamma$ is at least 6; in fact,
such values of $\gamma$ also ensure that the NRC-violating era is
quickly replaced by an inflationary period.

Very soon, then, $g\m{_c(\gamma,\,K,\,L_{INF})_{+---}}$ becomes
indistinguishable from the OVV metric \newline
$g\m{_{OVV}(2^{-\,(2/\gamma)}K,\,L_{INF})_{+---}}$ [equation
(\ref{eq:T})], which is, locally, the standard \emph{inflationary}
metric. Notice that one relic of the NRC-violating era survives:
the parameter K, which measures the radius of the torus at its
smallest. Thus K is some extremely small number.

This inflationary era ends in the conventional way at time
t$_{\m{EXIT}}$, and we switch [via some transitional geometry
which we shall not attempt to describe] to a metric like
$g\m{_s(4,\,K_{EXIT},\,L_{DE})_{+---}}$ [equation
(\ref{eq:EPSILON})] to describe the radiation-dominated era. Here
$\m{K_{EXIT}}$ is related to the size of the torus at time
t$_{\m{EXIT}}$, and $\m{L_{DE}}$ is the length scale appropriate
to the current Dark Energy phase --- that is, $\m{L_{DE}}$ is very
much larger than $\m{L_{INF}}$. At some time, before decoupling,
$g\m{_s(4,\,K_{EXIT},\,L_{DE})_{+---}}$ in its turn will be
replaced by a metric like $g\m{_s(3,\,K_{RAD},\,L_{DE})_{+---}}$,
where $\m{K_{RAD}}$ is related to the size of the torus at the end
of the radiation-dominated era. This is the metric appropriate to
the matter-dominated era. Finally,
$g\m{_s(3,\,K_{RAD},\,L_{DE})_{+---}}$ will eventually become
indistinguishable from yet another OVV metric, as the matter
dilutes and the Universe enters its final phase, in which the dark
energy dominates.

This picture is particularly natural if we are using the so-called
``low-scale" versions of inflation, in which the inflationary era
begins at a mass scale about 3 orders of magnitude below the
Planck mass. These arise naturally in string theory, but there are
difficulties in understanding the initial conditions for this kind
of inflation. However, Linde
\cite{kn:lindetypical}\cite{kn:lindenew} stresses that a simple
and elegant way to solve the problem of initial conditions for
low-scale inflation is to assume that the spatial sections of the
Universe are compact and have either zero or negative curvature.
As discussed in \cite{kn:reallyflat}, the negative case is not
acceptable in string theory because such spacetimes are unstable
in the Seiberg-Witten sense. Thus we are left with the flat case,
as discussed in this work. The flat, compact spatial sections
solve the problem provided that the system is able to sample an
entire spatial section. This will happen if the relevant part of
the Penrose diagram of the early Universe is much taller than it
is wide. Since any inflation-like expansion naturally expends
large amounts of conformal time, and since the parameter K is very
small, it should not be very difficult to arrange this. Thus the
whole picture is ideally suited to incorporate Linde's ideas,
though of course it must be made numerically much more precise; in
particular, one needs a quantum-gravitational account of the shape
of the Penrose diagram describing the earliest phases of cosmic
evolution.

A Penrose diagram illustrating the essential points of this
discussion is given in Figure 6.
\begin{figure}[!h]
\centering
\includegraphics[width=0.2\textwidth]{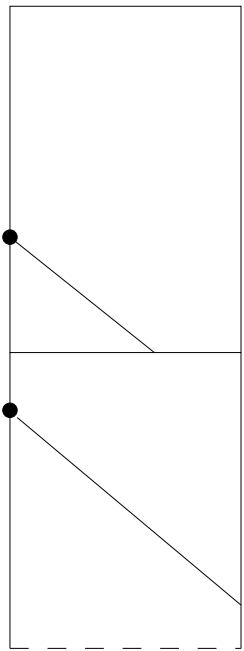}
\caption{Penrose diagram of the Toral Universe}
\end{figure}
The horizontal line in Figure 6 represents decoupling; the lower
dot represents a particle or gravitational perturbation in the
early Universe which has sampled an entire spatial section, as in
Linde's scenario; and the upper dot represents an observer like
ourselves, who is not \emph{yet} able to see an entire torus,
though he will be able to do so in future. There is much more to
be said about this picture, but let us turn to a more pressing
issue: what happens before t = 0, represented by the dashed line
at the bottom?

Here, again, we can seek guidance from Ooguri et al \cite{kn:OVV}.
In their discussion, the answer has two aspects, one Euclidean,
the other a mixture of Euclidean and Lorentzian.

In the Euclidean case, Ooguri et al note that the region of large
negative $\zeta$ in their metric given in (\ref{eq:P}) involves
circles of arbitrarily small radius. Moving towards still smaller
circles should be interpreted as motion towards \emph{larger}
circles in some T-dual theory. For example, Ooguri et al begin
with a IIB theory, so the dual description should be in terms of
IIA theory. In cosmological language, this should mean that all
physics on tori ``earlier" than t = 0 is completely equivalent to
dual physics on later tori: the past of t = 0 is equivalent to its
future, described in terms of different variables. In short, the
full Euclidean manifold describing our Universe in the OVV picture
consists of two identical ``trumpet-shaped" spaces joined together
smoothly at their narrowest point. The metric near this point
should resemble $g\m{_c(\gamma,\,K,\,L)_{++++}}$, for some large
value of $\gamma$.

This is in fact very similar to the standard way of describing the
manner in which \emph{string gas cosmologies} \cite{kn:brandvafa}
avoid being singular. Clearly it would be highly desirable to give
a precise description of the mapping of variables under T-duality
in cosmology, as Ooguri et al do for their manifold [in terms of
D3 branes on the IIB side and D2 branes on the IIA side].
Cosmology could actually implement string duality not just as a
mathematical equivalence, but in terms of spacetime geometry.

Ooguri et al are also interested in a possible interpretation in
terms of the Hartle-Hawking construction \cite{kn:hartle}. This
would involve regarding t = 0 as the section where there is a
transition from Euclidean geometry to Lorentzian. Recall that we
argued in Section 2 that \emph{both} ways of complexifying a
Euclidean space should have complementary roles in string
cosmology. If we take equation (\ref{eq:PI}) and complexify
\emph{time} instead of the spatial parameter K, we obtain a metric
of completely negative signature:
\begin{equation}\label{eq:VARSIGMA}
g\m{_c(\gamma,\,K,\,L)_{----} \;=\; -\,d\tau^2\; -\;
K^2\;cos^{(4/\gamma)}({{\gamma\,\tau}\over{2L}})\,[d\theta_1^2
\;+\; d\theta_2^2 \;+\; d\theta_3^2]}.
\end{equation}
Of course, such a metric is just an unusual way of describing a
Euclidean space; one has to take the absolute value before
computing a length, just as one does for spacelike displacements
in Minkowski space with signature ($+\;-\;-\;-$). [Note however
that a sphere has negative curvature in this signature.] Here
$\tau$ is a coordinate which takes its values in the interval
($-\pi$L/$\gamma$, 0]. The space is foliated by tori, which reach
a maximal radius of K at $\tau$ = 0. Hence, this Euclidean space
can be smoothly joined [at t = $\tau$ = 0] to the Lorentzian space
described by $g\m{_c(\gamma,\,K,\,L)_{+---}}$
[equation(\ref{eq:PI})]. It is significant that complexifying t in
$g\m{_c(\gamma,\,K,\,L)_{+---}}$ does not complexify the metric,
as would have happened if we had complexified time in
$g\m{_s(\epsilon,\,K,\,L)_{+---}}$ [equation (\ref{eq:EPSILON})]
for $\epsilon$ = 3 or 4. That is, the early period of NRC
violation reduces the extrinsic curvature of the spatial slices to
zero as the transition is approached from the Lorentzian side,
thus naturally allowing a smooth transition to a Euclidean regime
while removing the imaginary factor from the metric when time is
complexified. The two different forms of analytical continuation
used here are indeed complementary: one complexifies K in
$g\m{_s(\epsilon,\,K,\,L)_{+---}}$ to understand the physics of
the boundary, and one complexifies t in
$g\m{_c(\gamma,\,K,\,L)_{+---}}$ to understand the earliest phases
of this universe.

Thus, another possible answer to the question as to what happens
at t = 0 is that time ceases to have any meaning if we try to
probe beyond that point: the spacetime is replaced by the
Euclidean space with metric $g\m{_c(\gamma,\,K,\,L)_{----}}$. It
would be very interesting to attempt to construct the
corresponding wave function, in the manner described in
\cite{kn:OVV}. Note that $g\m{_c(\gamma,\,K,\,L)_{----}}$ is
singular at $\tau$ = $-\pi$L/$\gamma$, but it is not entirely
clear what this means. Note that, in the absence of a meaningful
``time" in the Euclidean region, there is no sense in which the
Universe can be said to ``evolve" from this point, so it may not
be as obnoxious as the more familiar kinds of cosmological
singularity. Nor does the singularity persist into the Lorentzian
regime, as happens in the case of the well-known Hawking-Turok
instanton \cite{kn:hawkturok}. Nevertheless one would like to know
whether there is some way of smoothing over this Euclidean
singularity, perhaps after the manner of \cite{kn:kirklin}. This
point remains to be investigated in detail.

 \addtocounter{section}{1}
\section*{\large{\textsf{7. Conclusion }}}

The main points of our discussion can be summarized as follows.
Ooguri, Vafa, and Verlinde have proposed a very unusual multiple
interpretation of the locally hyperbolic space H$^2/\bbz$, and
have explored the corresponding physics with a view to
interpreting a Lorentzian version as an accelerating cosmology. We
have extended this to the four-dimensional case, H$^4/\bbz^3$.
After clarifying the mathematics, we found that the cosmology
determined by this space is null geodesically incomplete. This
foreshadows the fact that any more realistic version of these
spacetimes is bound to be singular unless the Null Ricci Condition
is violated in the earliest Universe. [It is worth noting that,
however one proceeds, the presence of \emph{even one circle} in
the spacelike sections will lead, via the Andersson-Galloway
theorem, to the conclusion that the NRC has to be violated if
singularities are to be avoided. That is, almost any procedure of
the OVV kind is likely to lead to this conclusion.] We argued that
the NRC can however be violated without inducing instability, and
we proposed concrete ways in which the initial singularity can be
avoided.

Ooguri et al establish a kind of duality between an accelerating
cosmology, represented by H$^2/\bbz$, and a particular black hole
configuration. This is done by embedding H$^2/\bbz$ in string
theory, through a compactification of the form
(H$^2/\bbz)\,\times\,\m{S^2\,\times\,CY}$. The obvious analogue in
the case of H$^4/\bbz^3$, which has the local geometry of a space
of constant negative curvature, is a space of the form
(H$^4/\bbz^3)\,\times\,$FR, where FR denotes a Freund-Rubin space.
Freund-Rubin compactifications have recently been revived
\cite{kn:bobby}\cite{kn:lambert}\cite{kn:valandro} and it would be
interesting to study OVV cosmology from that point of view. In
particular, one should study the ``OVV dual" of the Freund-Rubin
embedding of our cosmologies. Note that Freund-Rubin
compactifications themselves have a dual interpretation in terms
of three-dimensional conformal field theories \cite{kn:spence}.

One striking feature of our analysis has been the remarkable way
in which the introduction of matter affects the OVV spacetime,
rendering it non-perturbatively stable in string theory when the
NRC is satisfied [and even --- sometimes --- when it is not]. Now
Firouzjahi et al \cite{kn:tye} have argued that the introduction
of matter into de Sitter space forces a modification of the
Hartle-Hawking wave function [as also do higher metric modes].
Clearly it would be desirable to reconsider the findings of Ooguri
et al from this point of view. The spacetimes we have considered
here provide a concrete background for such an investigation.

We have stressed that the OVV construction introduces a new length
scale when H$^2$ or H$^4$ is partially
compactified\footnote{Strictly speaking, there are actually three
such parameters in the H$^4$ case.}; this is the parameter K in
$g\m{_s(\epsilon,\,K,\,L)_{+---}}$ [which determines the K in
$g\m{_c(\gamma,\,K,\,L)_{+---}}$ when the latter is used in the
way we have suggested]. The ratio L/K determines the shape of the
Penrose diagrams for these spacetimes, and this shape is of
profound physical significance \cite{kn:challenge}. One would
certainly hope that it is fixed by the wave function of the
Universe, and proving that the wave function is peaked at a
physically reasonable value of L/K is an important challenge for
these theories.

 \addtocounter{section}{1}
\section*{\textsf{Acknowledgements}}
The author is extremely grateful to Wanmei for preparing the
diagrams and for moral support. He also sincerely thanks all those
who have made possible his visit to the High Energy, Cosmology,
and Astroparticle Physics Section of the Abdus Salam International
Centre for Theoretical Physics, where the work described here was
done.

\addtocounter{section}{1}
\section*{\textsf{Appendix: About The Andersson-Galloway Theorem}}
In this appendix we briefly explain the terminology used by
Andersson and Galloway \cite{kn:andergall}\cite{kn:gall}, and
comment on the conditions assumed in their theorem used above.
This is important, because our argument is based on the claim that
there is only one way to circumvent the theorem --- to violate the
Null Ricci Condition.

A four-dimensional spacetime M$_4$ with Lorentzian metric
$g_{\mathrm{M}}$ is said to have a {\em regular future spacelike
conformal boundary} if M$_4$ can be regarded as the interior of a
spacetime-with-boundary X$_4$, with a [non-degenerate] metric
$g_{\mathrm{X}}$ such that the boundary is {\em spacelike} and
lies to the future of all points in M$_4$, while $g_{\mathrm{X}}$
is conformal to $g_M$, that is, $g_{\mathrm{X}}$ =
$\Omega^2g_{\mathrm{M}}$, where $\Omega$ = 0 along the boundary
but d$\Omega \neq 0$ there. This is just a technical formulation
of the idea that the  Penrose completion should be well-behaved.
There are examples of accelerating spacetimes which do not have a
regular future spacelike conformal boundary --- the Nariai
spacetime \cite{kn:rp3} is of this type
--- but these spacetimes typically violate the NEC [\emph{some}
``phantom"
cosmologies, though certainly not all \cite{kn:smash}, are
singular towards the future, with the singularity at finite proper
time]
 or are highly non-generic, like Nariai spacetime [which is on the
 very brink of having a naked singularity]. One certainly
should not hope to escape from the conclusions of the
Andersson-Galloway theorem by resorting to such examples.

A spacetime is said to be \emph{globally hyperbolic} if it
possesses a Cauchy surface, that is, a surface on which data can
be prescribed which determine all physical fields at later and
earlier events in spacetime, since all inextensible timelike and
null curves intersect it. This forbids naked singularities, but it
also disallows ordinary AdS. The dependence of the
Andersson-Galloway theorem on this assumption might lead one to
ask whether our conclusions can be circumvented by dropping global
hyperbolicity. This is a topical suggestion, since it has recently
been claimed \cite{kn:veronika} that string theory allows
spacetimes which violate global hyperbolicity even more
drastically than AdS. To see why this too will not work here, we
need the following definition.

A spacetime with a regular future spacelike conformal completion
is said to be {\em future asymptotically simple} if every future
inextensible null geodesic has an endpoint on future conformal
infinity. This just means that there are no singularities to the
future --- obviously a reasonable condition to impose in our case,
since it would be bizarre to suppose that singularities to the
future can somehow allow us to avoid a Big Bang singularity.
Andersson and Galloway \cite{kn:andergall} show, however, that if
a spacetime has a regular future spacelike conformal completion
and is future asymptotically simple, \emph{then it has to be
globally hyperbolic}. Thus it would not be reasonable to drop this
condition in our context. Notice that this discussion has a more
general application: it means that, if the future of our Universe
resembles that of de Sitter spacetime, any attempt to violate
global hyperbolicity will necessarily cause a singularity to
develop to the future. It would be interesting to understand this
in the context of \cite{kn:veronika}.

Finally, one might wonder whether a compact flat three-manifold
can in fact have a first homology group which is pure torsion. The
rather surprising answer is that it can: there is a unique
manifold of this kind, the \emph{didicosm}, described in
\cite{kn:conway}\cite{kn:reallyflat}. The didicosm has the form
T$^3/[\bbz_2\;\times\;\bbz_2$], that is, it is a quotient of the
three-torus. However, if it were possible to construct a
singularity-free spacetime metric on
$\bbr\,\times\,\m{T}^3/[\bbz_2\;\times\;\bbz_2]$ without violating
the NRC, this metric would pull back, via an obvious extension of
the covering map T$^3 \rightarrow
\m{T}^3/[\bbz_2\;\times\;\bbz_2$], to a non-singular metric on
$\bbr\,\times\,\m{T}^3$, also satisfying the NRC. This is a
contradiction. Similarly, of course, one cannot escape the
conclusions of the theorem by allowing spacelike future infinity
to be non-orientable.

We conclude that the only physically reasonable way to avoid a
Bang singularity in a spacetime of OVV topology is indeed to
violate the NRC.

\end{document}